\documentclass[traditabstract]{aa}
\usepackage[pdftex]{graphicx}
\usepackage{txfonts}
\usepackage{color}
\usepackage{natbib}
\usepackage{epsfig}
\usepackage{hyperref}
\usepackage{sidecap}

\begin{document}


\title{Simultaneous EUV and radio observations of bidirectional plasmoids ejection during magnetic reconnection}
\author{Pankaj Kumar \and K.S. Cho }
\institute{Korea Astronomy and Space Science Institute (KASI), Daejeon, 305-348, Republic of Korea \\
\email{pankaj@kasi.re.kr}}

\abstract
{We present a multiwavelength study of the X-class flare, which occurred in active region (AR) NOAA 11339 on 3 November 2011. The extreme ultraviolet (EUV) images recorded by SDO/AIA show the activation of a remote filament (located north of the AR) with footpoint brightenings about 50 min prior to the flare occurrence. The kinked filament rises-up slowly and after reaching a projected height of $\sim$49 Mm, it bends and falls freely near the AR, where the X-class flare was triggered. Dynamic radio spectrum from the Green Bank Solar Radio Burst Spectrometer (GBSRBS) shows simultaneous detection of both positive and negative drifting pulsating structures (DPSs) in the decimetric radio frequencies (500-1200 MHz) during the impulsive phase of the flare. The global negative DPSs in solar flares are generally interpreted as a signature of electron acceleration related to the upward moving plasmoids in the solar corona. The EUV images from AIA 94 \AA~ reveal the ejection of multiple plasmoids, which move simultaneously upward and downward in the corona during the magnetic reconnection. The estimated speeds of the upward and downward moving plasmoids are $\sim$152-362 and $\sim$83-254 km s$^{-1}$, respectively. These observations strongly support the recent numerical simulations of the formation and interaction of multiple plasmoids due to tearing of the current-sheet structure. On the basis of our analysis, we suggest that the simultaneous detection of both the negative and positive DPSs is most likely generated by the interaction/coalescence of the multiple plasmoids moving upward and downward along the current-sheet structure during the magnetic reconnection process. Moreover, the differential emission measure (DEM) analysis of the active region reveals presence of a hot flux-rope structure (visible in AIA 131 and 94 \AA) prior to the flare initiation and ejection of the multi-temperature plasmoids during the flare impulsive phase.}          

\authorrunning{Kumar \& Cho}
\titlerunning{Bidirectional plasmoids ejection during magnetic reconnection}

\maketitle
\section{INTRODUCTION}
It is well known that magnetic reconnection plays a key role in the release of magnetic energy stored in sheared/twisted magnetic fields, leading to the onset of solar flares and coronal mass ejections (CMEs) \citep{priest2000,asc2004}. In the magnetic
reconnection process, oppositely directed field lines break-up and rejoin resulting 
in the release of magnetic energy into thermal energy and particle 
acceleration (e.g., \citealt{sweet1958}, \citealt{parker1963}, \citealt{petschek1964}). The standard model of solar flare, known
as the CSHKP model, explains the energy release process \citep{carm1964,stur1966,hirayama1974,kopp1976}, which is supported by several observational findings, e.g., 
cusp-shaped loops \citep{tsuneta1992}, inflows and outflows \citep{yokoyama2001,savage2010,savage2012,takasao2012}, downflow signatures \citep{mckenzie2000,mckenzie2001,innes2003,asai2004,savage2011}, plasmoid ejections \citep{shibata1995}, loop-top hard X-ray sources \citep{masuda1994,sui2003}, X-ray jets \citep{shibata1992,shimojo1996}, bidirectional-jets \citep{innes1997}, and flux rope/loop interactions \citep{kumar2010a,kumar2010b,kumar2010c,torok2011}. According to the CSHKP model, the magnetic reconnection takes place in a vertical current-sheet located above an underlying closed loop system and the filament/prominence eruption plays a key role in the triggering of fast reconnection \citep{forbes2000,lin2000,kumar2012a}.

Although there are some fundamental problems in the theory of magnetic reconnection. For example, in the Sweet-Parker reconnection model \citep{sweet1958,parker1957}, the reconnection-rate is too slow (10$^{-4}$-10$^{-6}$) to explain the magnetic energy release in solar flares. The \citet{petschek1964} reconnection model predicts a faster reconnection rate (i.e., 0.01-0.1), considering slow-mode MHD shocks in the outflow region. This model explains the observed reconnection rate in most of the solar flares. However, observational value of the reconnection rate varies from $\sim$0.2-0.001 \citep{isobe2005,narukage2006,takasao2012}. MHD simulations have revealed that the localized resistivity can assist to obtain the fast reconnection rate (i.e., close to unity) \citep{ugai1977,yokoyama1994}. When the diffusion region becomes long enough (similar to the Sweet-Parker model), the tearing mode instability can cause the onset of bursty reconnection, which involves the formation of several magnetic islands (i.e., plasmoids) in the current sheet \citep{furth1963,kliem1995,priest2000,shibata2011}.

\citet{shibata1995} and \citet{shibata2001} extended the CSHKP model by unifying reconnection and plasmoid ejection, and emphasized on the importance of plasmoid ejection in the reconnection process, which is known as ``plasmoid-induced-reconnection" model.
In this model, the bursty reconnection is triggered by a plasmoid ejection that leads to build-up of the magnetic energy in a vertical current-sheet. When a plasmoid is ejected, the inflow is induced due to the conservation of mass, which results in the enhancement of the reconnection rate. A plasmoid formed above the current sheet can also accelerated due to the faster reconnection outflow. \citet{nishida2009} performed MHD simulations of the solar flares using different values of resistivity and plasmoid velocity, and found that the reconnection rate correlates positively with the plasmoid velocity. Therefore, the plasmoid ejection plays a key role in triggering fast magnetic reconnection and is an observational evidence of magnetic reconnection in a solar flare. For example, in the
 soft X-ray and EUV images, upward-moving hot plasma blobs are frequently observed \citep{shibata1995,ohyama1998,kim2005,kumar2013a,kumar2013} and the white-light coronagraph observations often show rising blob-like features in the wake of CMEs \citep{ko2003,lin2005,liu2011}. 
   The typical size of the plasmoid varies from $\sim$10$^{9}$ cm (in compact flares) to $\sim$10$^{11}$ cm (in CMEs) and their speed ranges between $\sim$10-1000 km s$^{-1}$ \citep{shibata2001}. Moreover, the size and velocity of multiple plasmoids formed during the the bursty reconnection (due to tearing instability) varies from $\sim$3-4$\arcsec$ and $\sim$89-460 km s$^{-1}$, respectively \citep{shen2011,takasao2012}. On the basis of numerical simulation by \citet{samtaney2009}, the number of plasmoids (and spatial scales) formed due to the tearing-mode instability, depends on the Lundquist number (i.e., S$^\frac{3}{8}$).

Drifting-pulsating structures (DPSs) are the slowly drifting radio spectrogram features of short durations ($\sim$1-3 min), that consists of several quasi-periodic pulsations. They are generally observed in the decimetric frequency range, (e.g., $\sim$0.6-2.0 GHz) \citep{khan2002}. Their mean bandwidth is about 500 MHz, and the frequency drift ranges from -20 to -5 MHz s$^{-1}$ \citep{karlicky2012}.
Based on a 2D numerical MHD simulation, \citet{kliem2000} proposed that the DPSs are generated during a bursty magnetic-reconnection when interacting plasmoids are formed. They proposed that the electrons are accelerated during these processes and trapped in the plasmoids that generate individual pulses of the observed DPSs. \citet{karl2004} used the concept of fractal reconnection proposed by \citet{shibata2001} and considered several plasmoids to explain simultaneous observation of several DPSs. The interaction of plasmoids is also revealed in the numerical simulations and negative/positive DPSs corresponds to upward/downward moving plasmoids exciting the radio emissions within progressively lower/higher 
 density regions \citep{barta2007,karl2007,ning2007,barta2008}. According to \citet{barta2008}, the upward and downward motion of the plasmoid is determined by the magnetic reconnection rate above and below the plasmoid. If the reconnection rate above the plasmoid is higher/lower than below, then the plasmoid moves downward/upward. The interaction of downward moving plasmoids with the flare loop arcade or loop-top kernel are also noticed in the simulation and observation \citep{barta2007,riley2007,kolo2007}. \citet{milligan2010} observed downward moving plasmoid (12 km s$^{-1}$) interacting with flare loop-top and showed the enhanced nonthermal emission in the corona at the time of the merging, suggesting the additional particle acceleration. Therefore, high resolution observations are extremely important to support the above simulation results.

So far, there is no clear simultaneous observation of both the positive and negative DPSs associated with the multiple plasmoid ejections in EUV or X-ray images. Thanks to the SDO/AIA high-resolution observations, which make it possible to observe this rare phenomena for the first time in an X-class flare occurred on 3 November 2011. In this paper, we present the simultaneous EUV and radio observational evidences of the bidirectional plasmoid ejections along the current-sheet structure during magnetic reconnection. 
In section 2, we present the multiwavelength observational data analysis and in the last section, we discuss our results.


\begin{figure}
\centering{
\includegraphics[width=9cm]{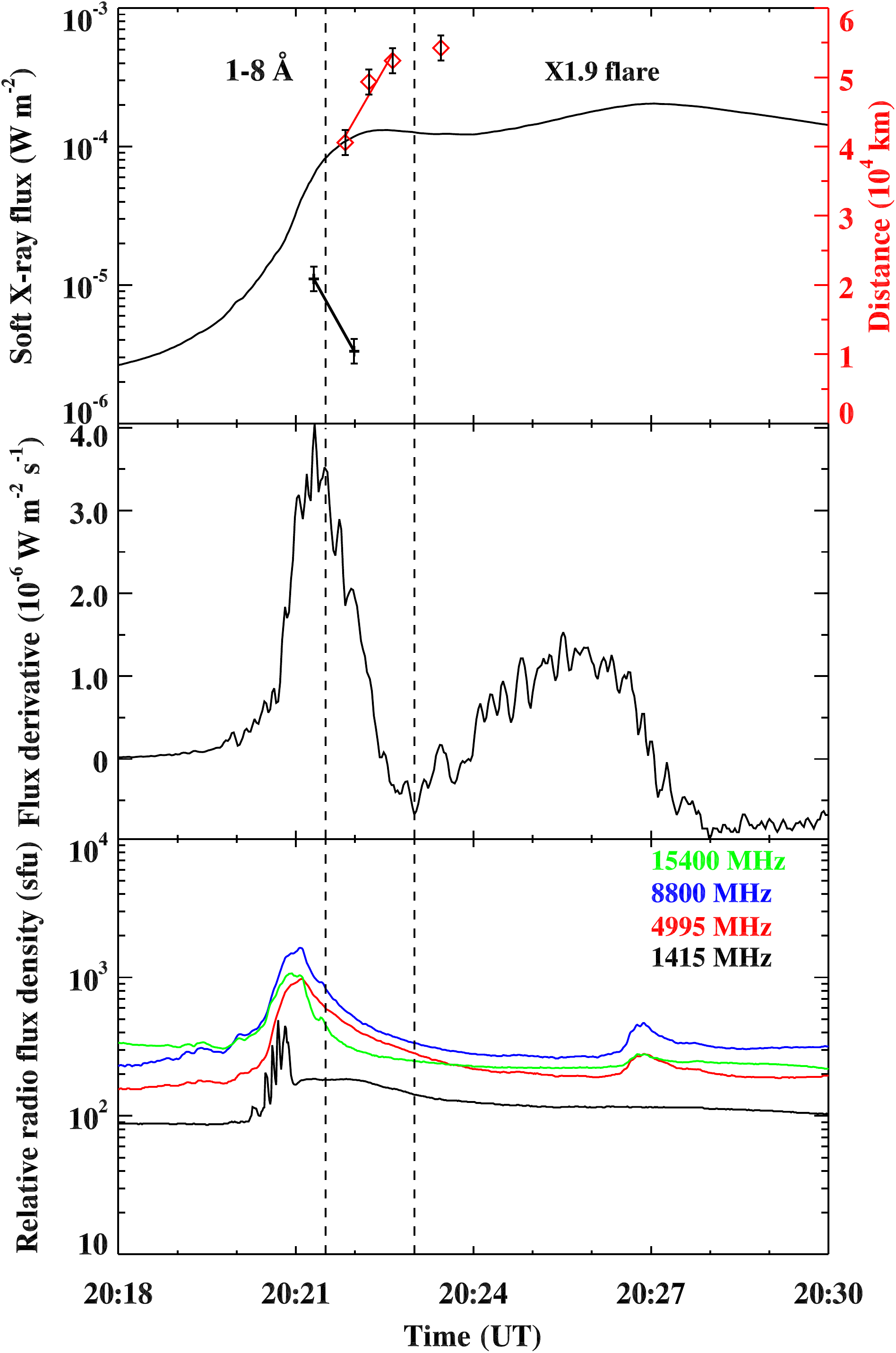}
}
\caption{Top: GOES soft X-ray flux in the 1-8 \AA~ channel and distance-time measurements of the upward and downward moving plasmoids.  Middle: GOES soft X-ray flux derivative. Bottom: RSTN 1 sec cadence radio flux profiles in 1445, 4995, 8800, and 15400 MHz frequencies during the flare on 3 November 2011. The two vertical dotted lines indicate the timings of the observed DPSs.}
\label{flux}
\end{figure}
\begin{figure*}
\centering{
\includegraphics[width=5cm]{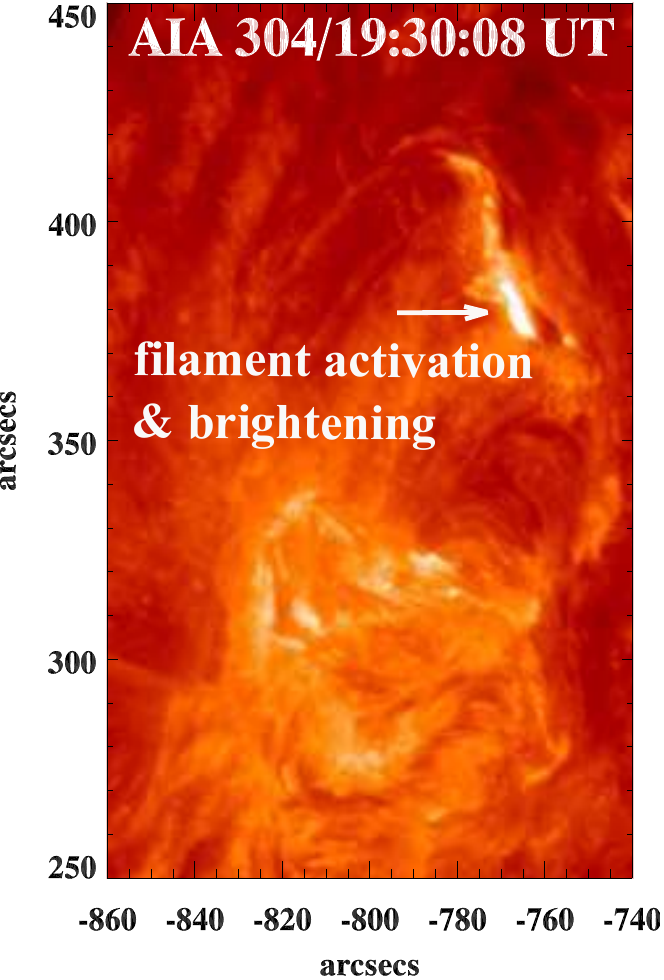}
\includegraphics[width=5cm]{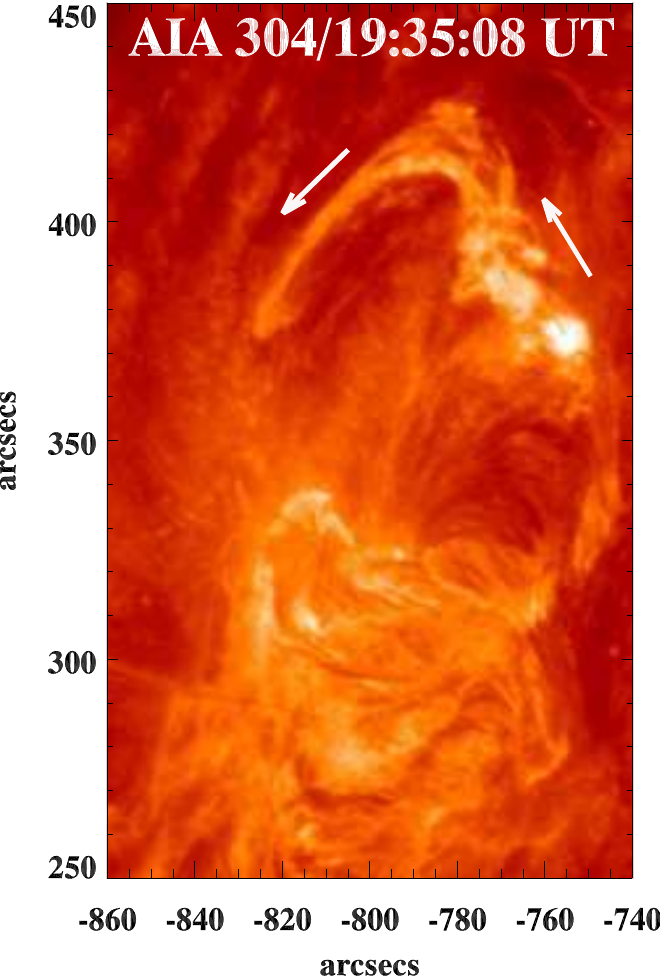}
\includegraphics[width=5cm]{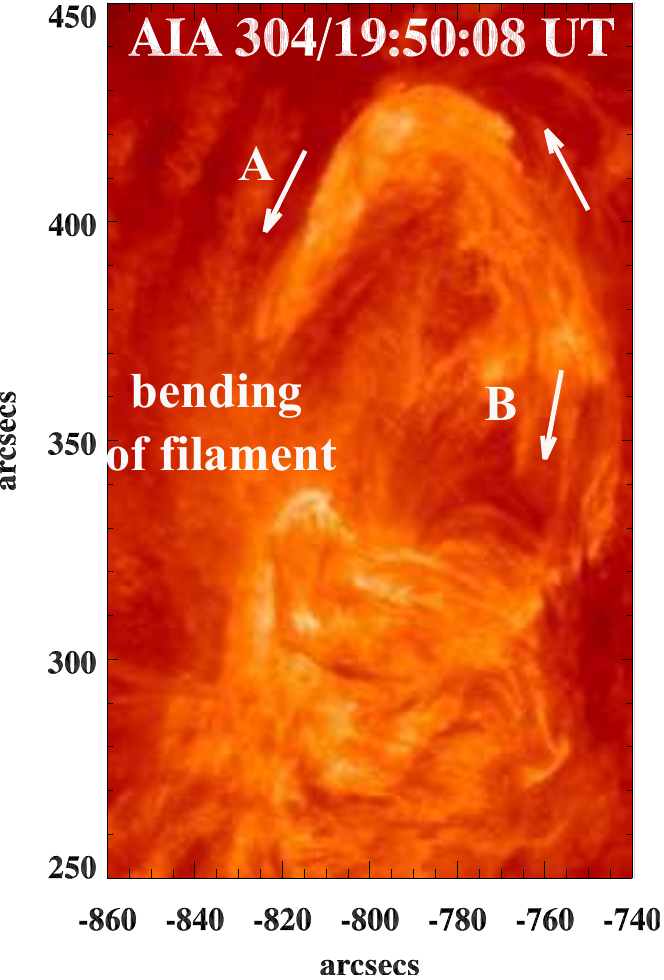}
}
\centering{
\includegraphics[width=5cm]{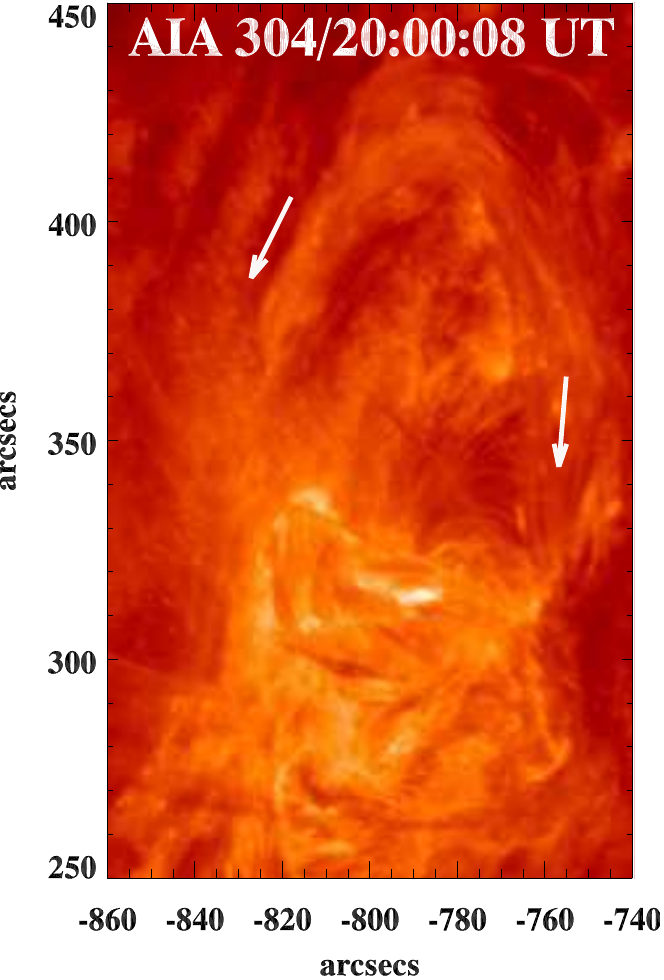}
\includegraphics[width=5cm]{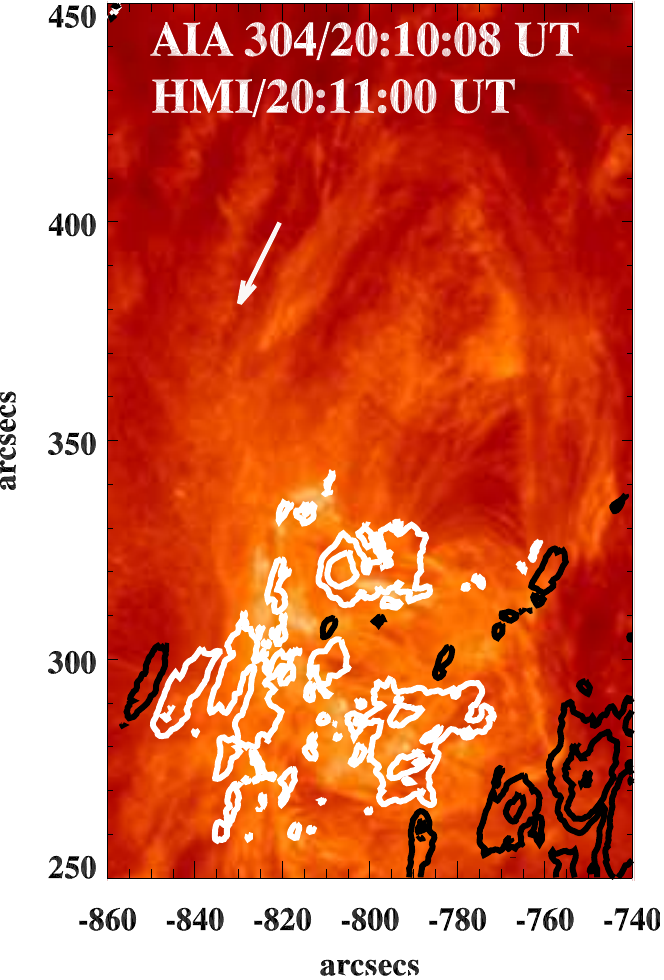}
\includegraphics[width=5cm]{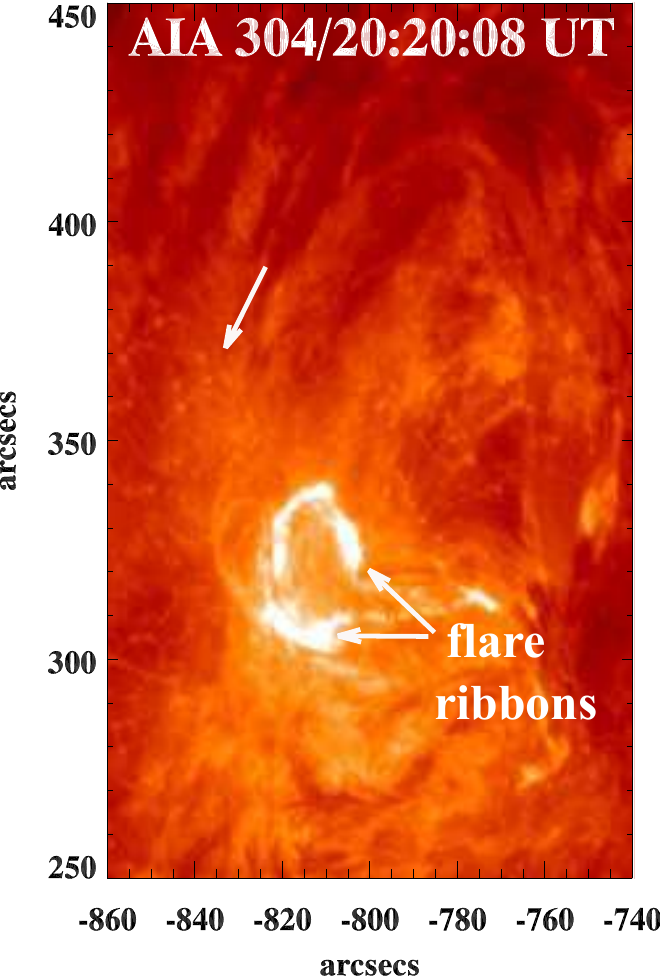}
}
\centering{
\includegraphics[width=5cm]{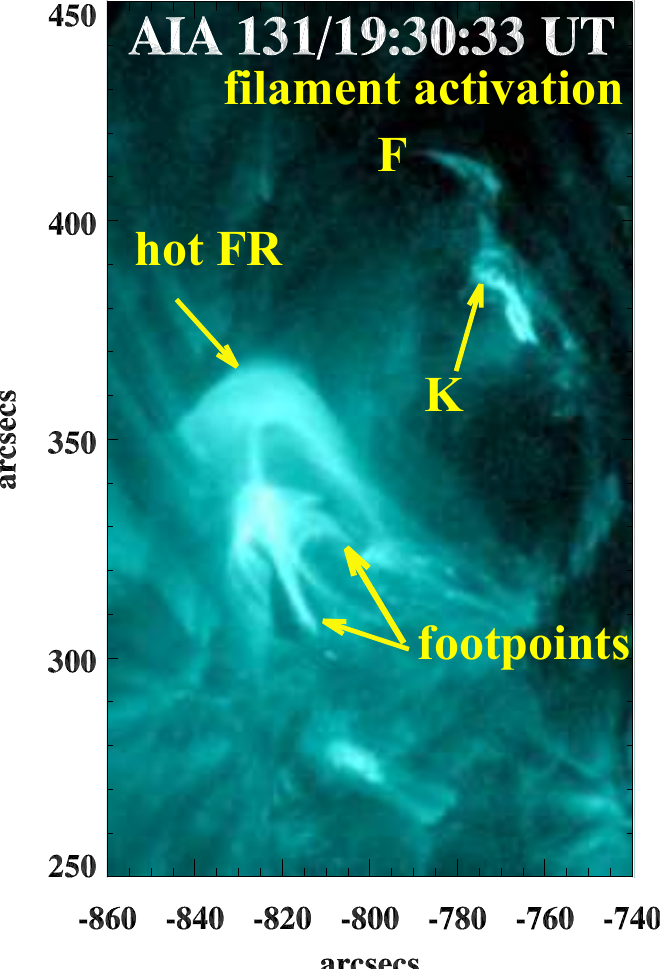}
\includegraphics[width=5cm]{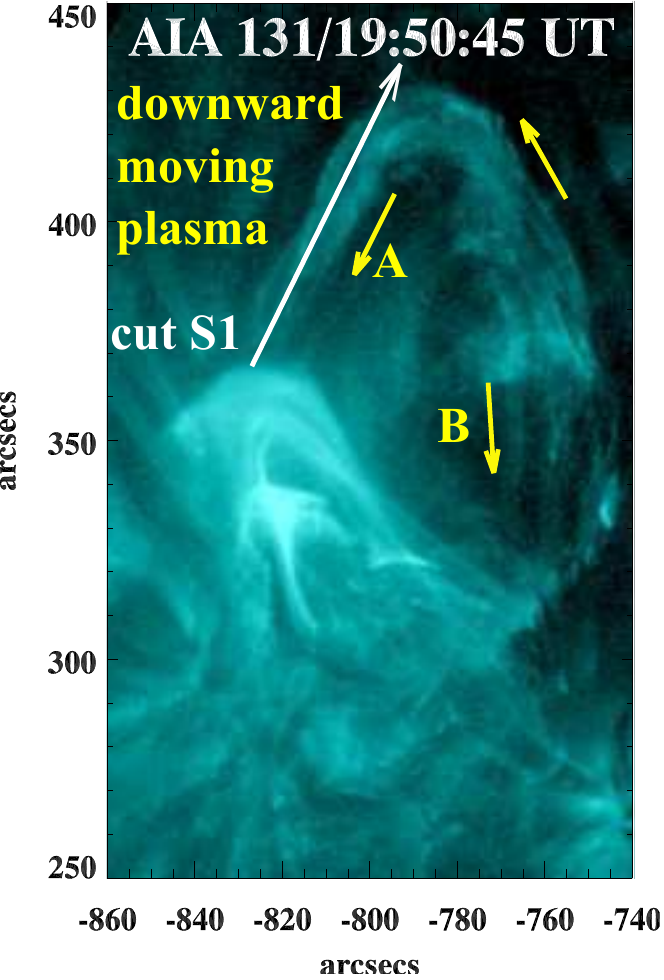}
\includegraphics[width=5cm]{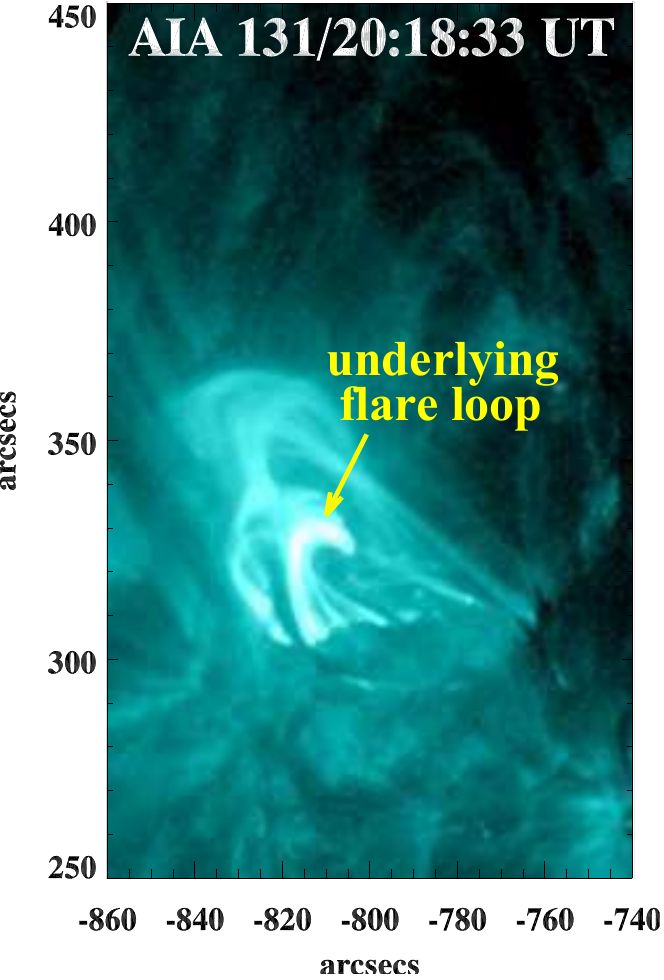}
}
\caption{SDO/AIA EUV images at 304 \AA \ (top and middle) and 131 \AA~(bottom) showing activation and bending of the filament toward the AR hot flux-rope structure on 3 November 2011. AIA 304 \AA~ image at 20:10:08 UT is overlaid by the HMI magnetogram contours of positive (white) and negative (black) polarities. The contour levels are $\pm$500,$\pm$1000, and $\pm$2000 G. The temporal evolution in the 304 \AA~ and 131 \AA~ bands over about one hour until the flare sets in can be found in a movie available in the online edition.}
\label{aia304}
\end{figure*}


\begin{figure*}
\centering
{
\includegraphics[width=14cm]{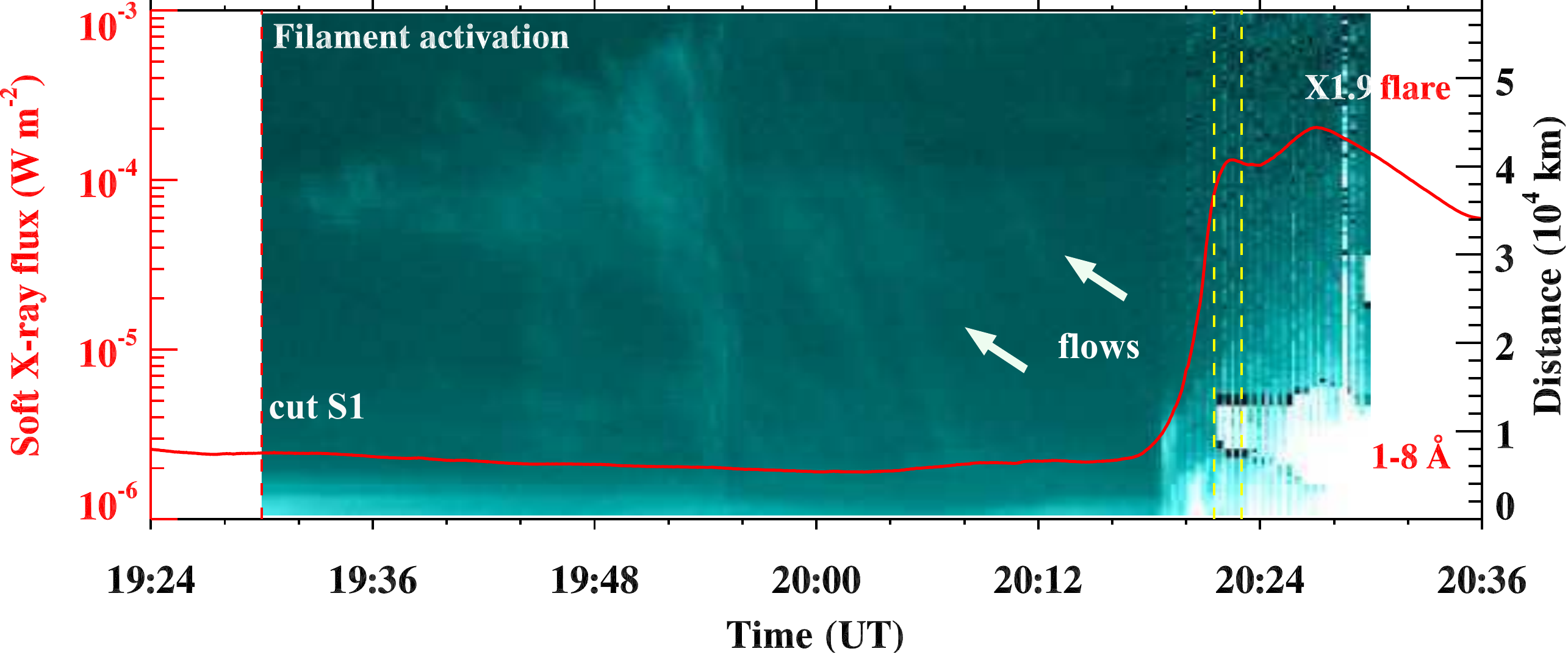}
}
\caption{GOES soft X-ray flux profile in the 1-8 \AA~ channel and space-time plot along slice `S1' of the AIA 131 \AA~ images on 3 November 2011. Two vertical dashed lines (yellow) indicate the timing of multiple plasmoid ejections during the impulsive phase of the flare. The timing of DPSs and plasmoid ejection is consistent with each other (see Figure 1).}
\label{sl131}
\end{figure*}

\section{Observations and results}

The Atmospheric Imaging Assembly (AIA) instrument onboard the Solar Dynamics Observatory (SDO) mission observes full disk images of the Sun at a spatial resolution of 1.2$^{\prime\prime}$ and its field of view covers $\sim$1.3 R$_\odot$. In this study, we used AIA images observed in 171 \AA~ (Fe IX, T$\approx$0.7 MK), 131 \AA~ (Fe VIII/XXI, T$\approx$0.4 \& 11 MK), 94 \AA~ (Fe XVIII, T$\approx$7 MK), and 304 \AA~ (He II, T$\approx$0.05 MK). The AIA images observed at 12 sec cadence cover chromospheric to coronal heights \citep{lemen2012}, which are useful to study the evolutionary aspects of the X-class flare. We also used the Helioseismic and Magnetic Imager (HMI) magnetogram to study the magnetic field configuration at the flare site \citep{schou2012,scherrer2012}.

\begin{figure*}
\centering{
\includegraphics[width=4.4cm]{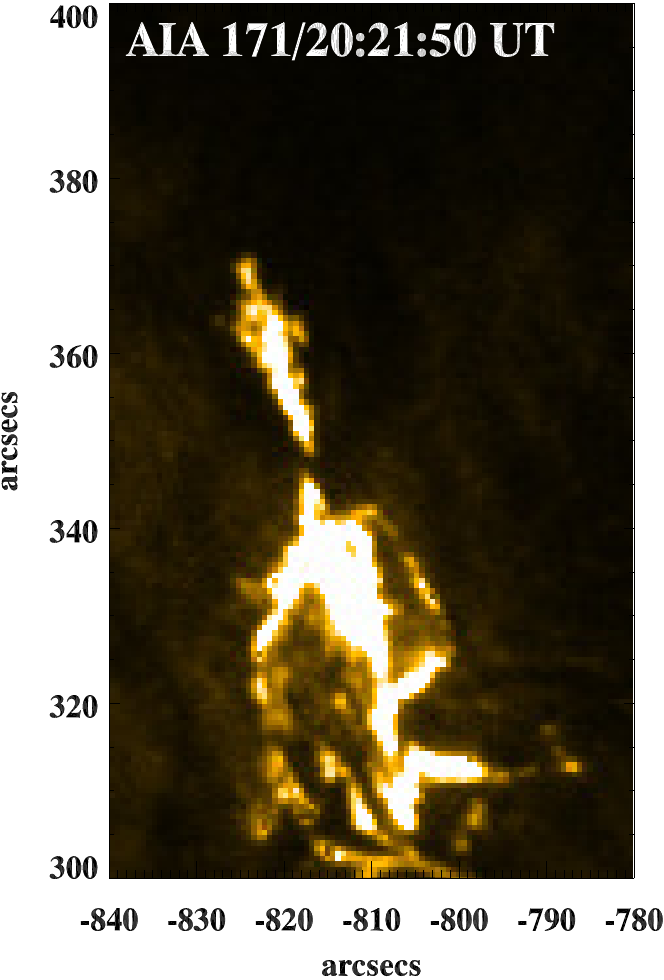}
\includegraphics[width=4.4cm]{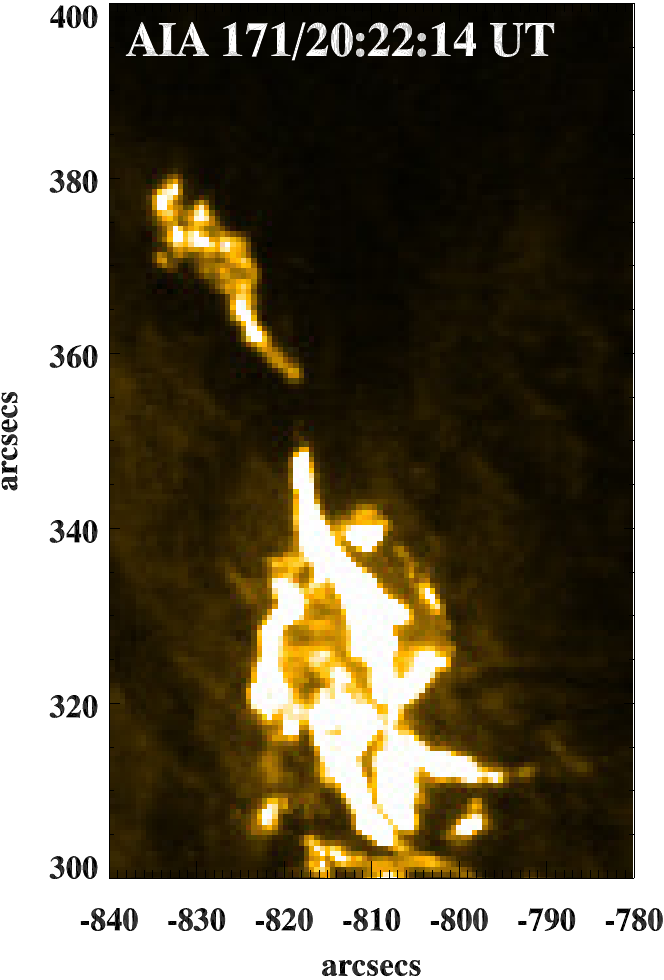}
\includegraphics[width=4.4cm]{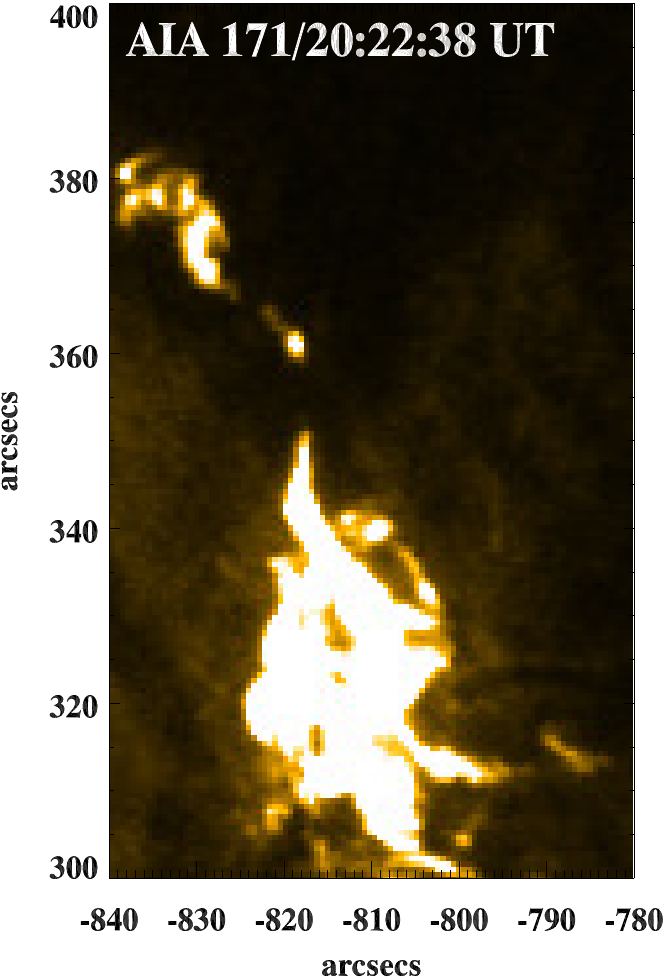}
}

\centering{
\includegraphics[width=4.4cm]{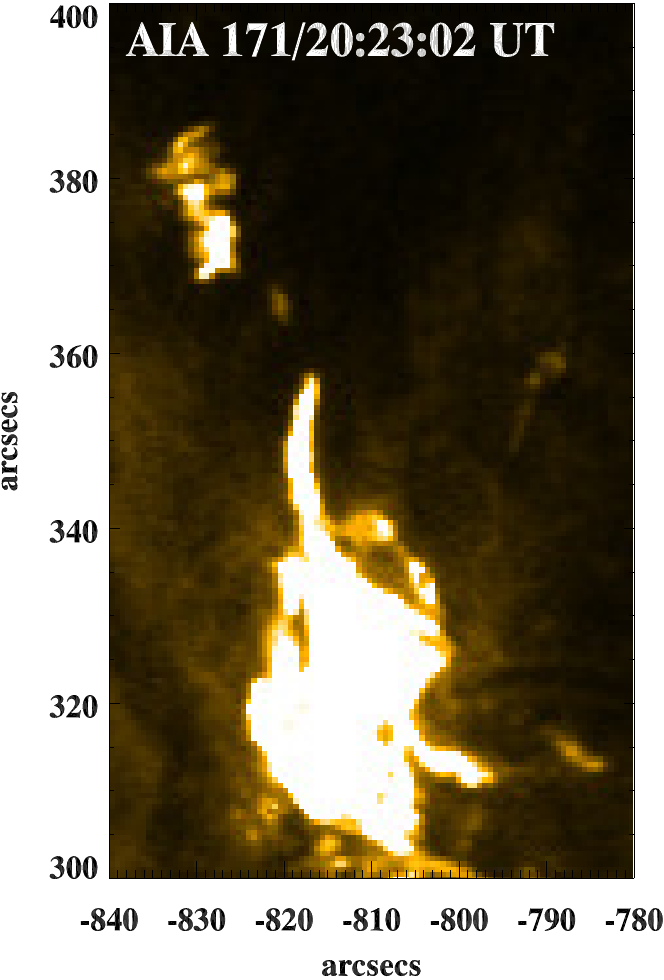}
\includegraphics[width=4.4cm]{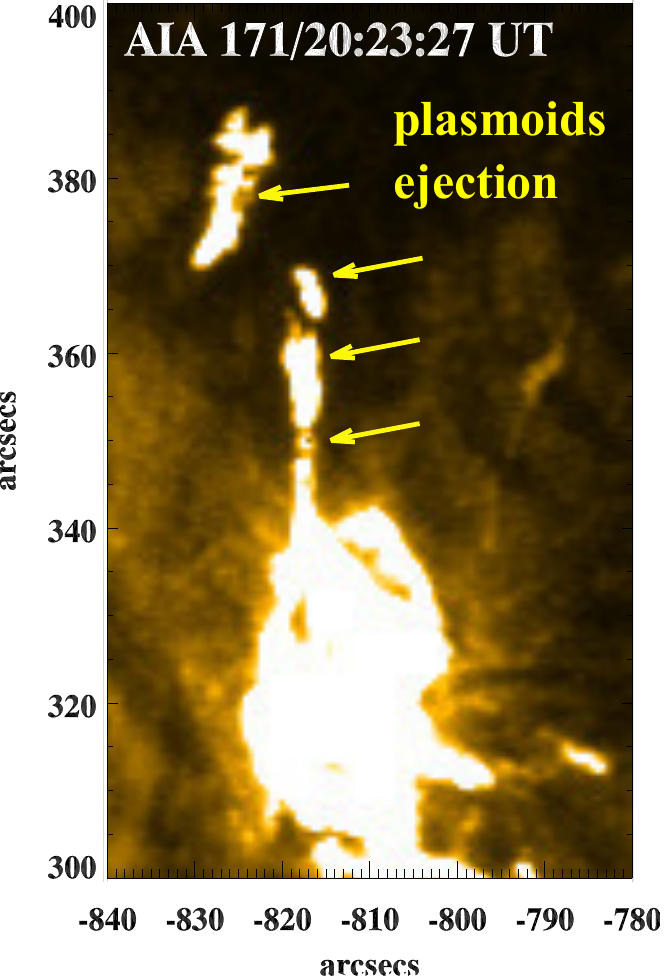}
\includegraphics[width=4.4cm]{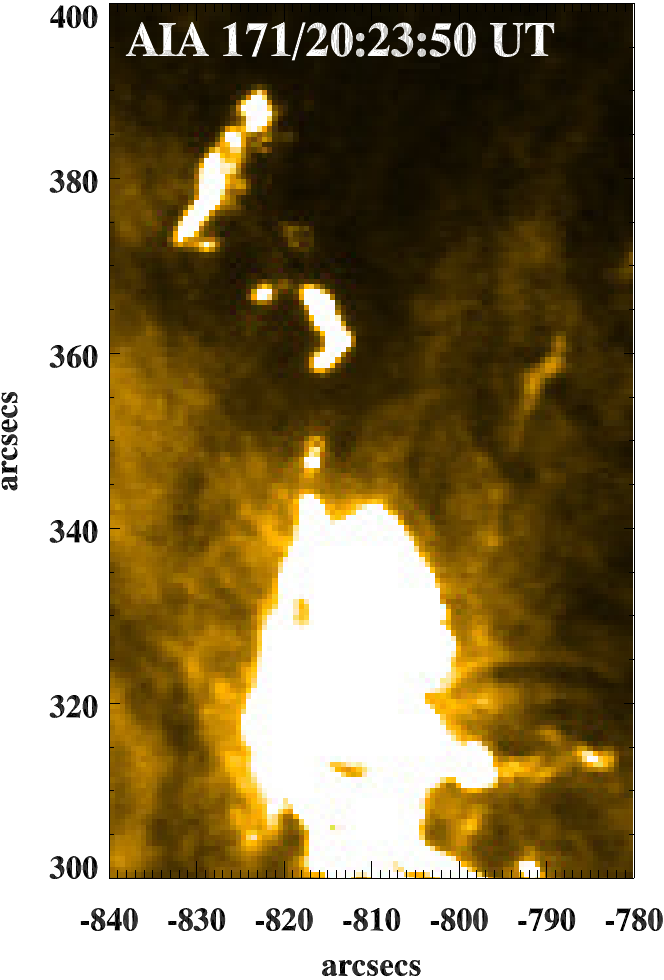}
}

\centering{
\includegraphics[width=4.4cm]{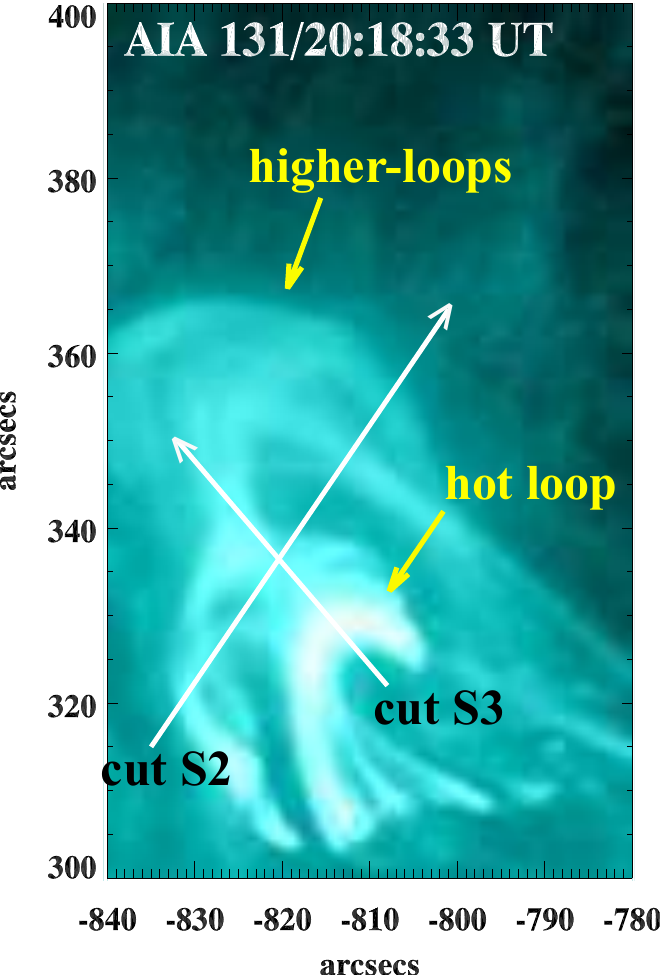}
\includegraphics[width=4.4cm]{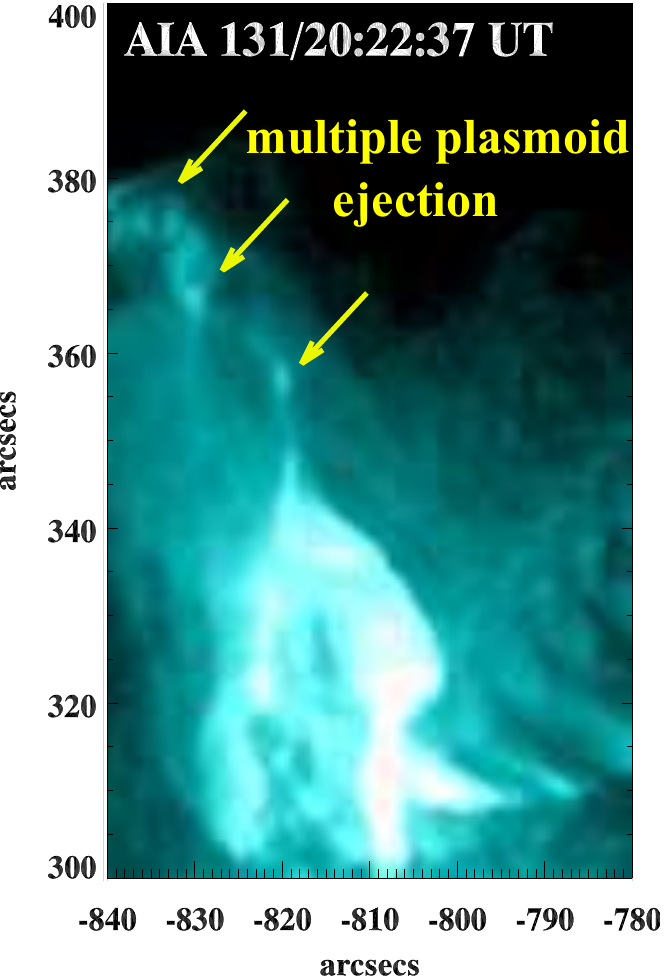}
\includegraphics[width=4.4cm]{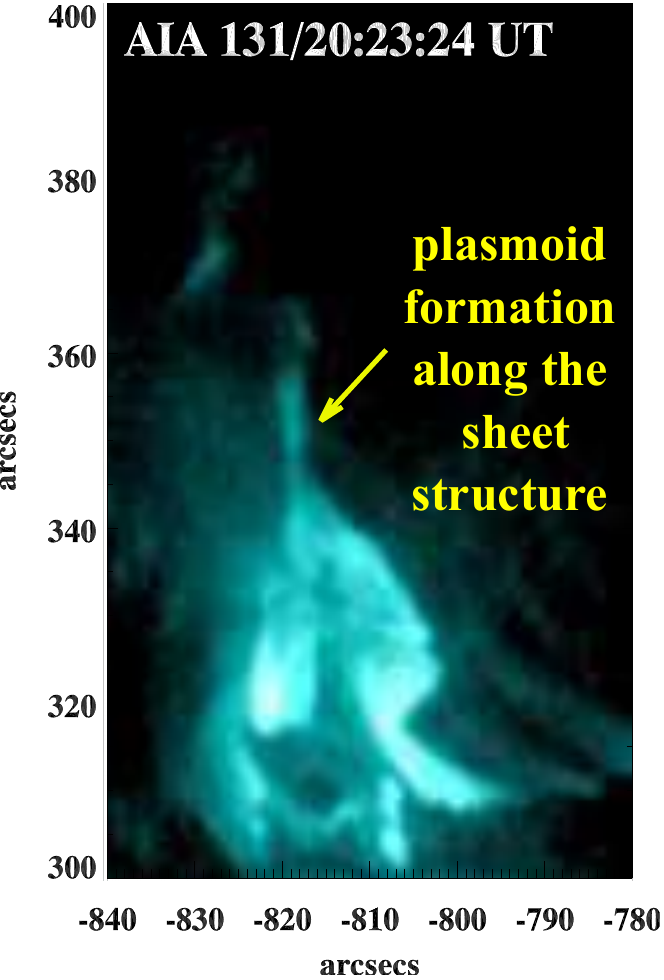}
}

\centering{
\includegraphics[width=4.4cm]{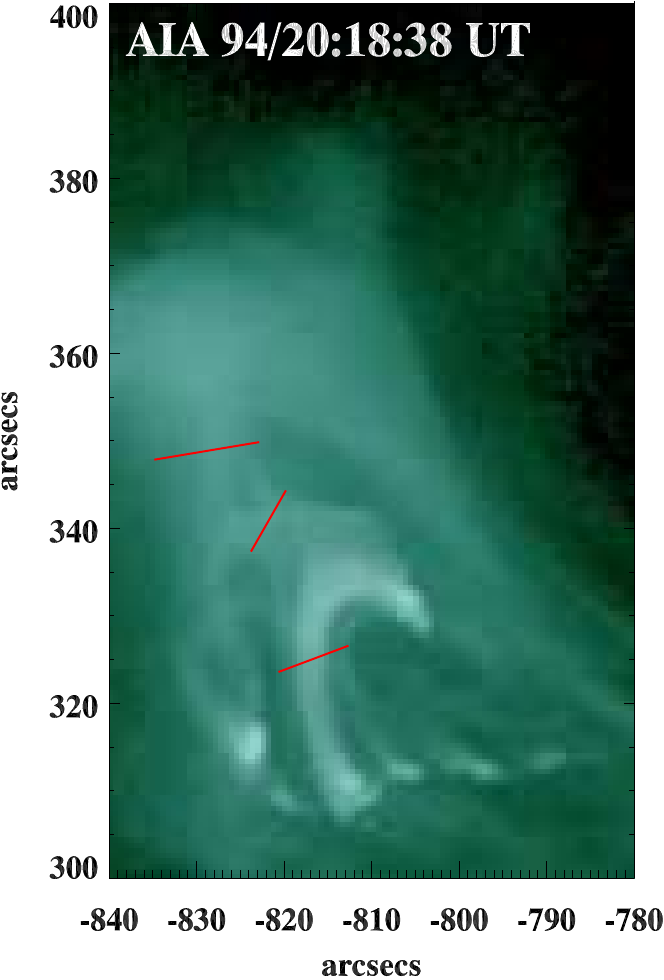}
\includegraphics[width=4.4cm]{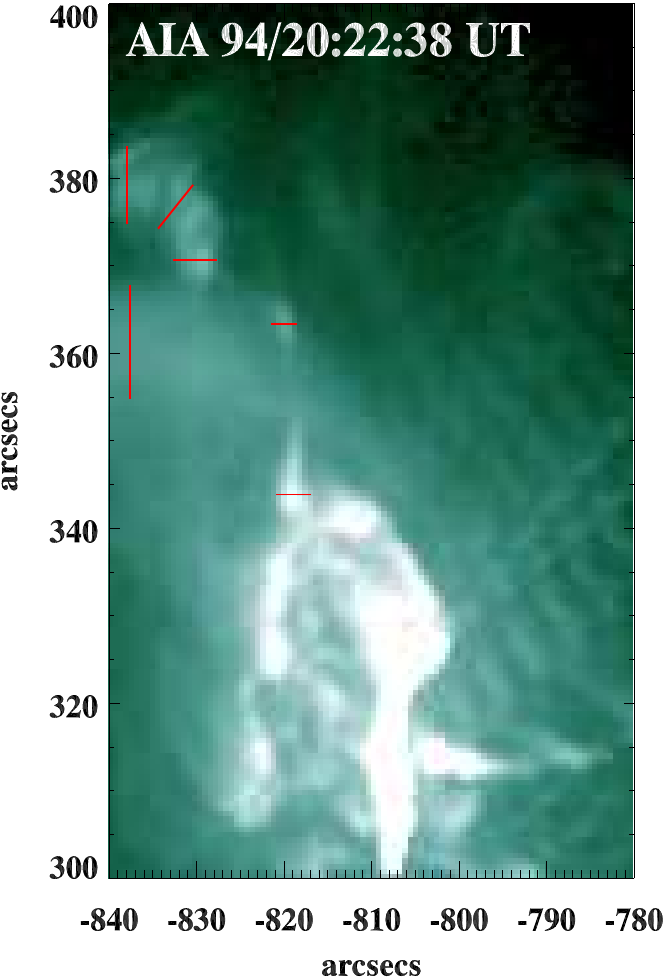}
\includegraphics[width=4.4cm]{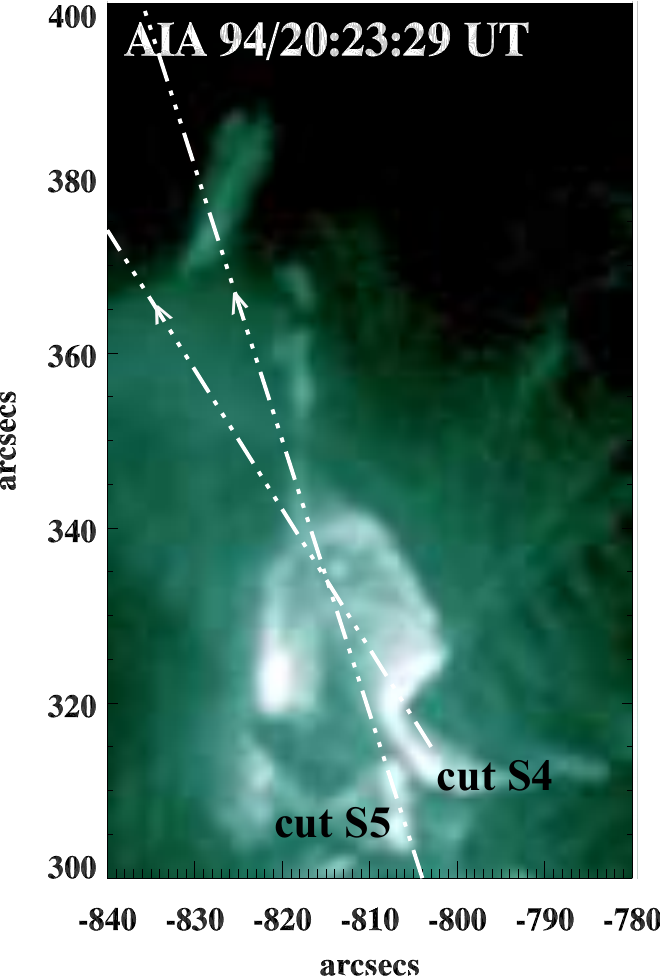}
}

\caption{SDO/AIA EUV images at 171 (top two rows), 131 (middle), and 94 (bottom) \AA~ showing multiple plasmoid ejections during the X-class flare on 3 November 2011. Red lines in the AIA 94 \AA~ images (bottom) indicate the approximate thickness of the structures which is used for the density estimation.}
\label{aia171}
\end{figure*}

\begin{figure*}
\centering{
\includegraphics[width=9.3cm]{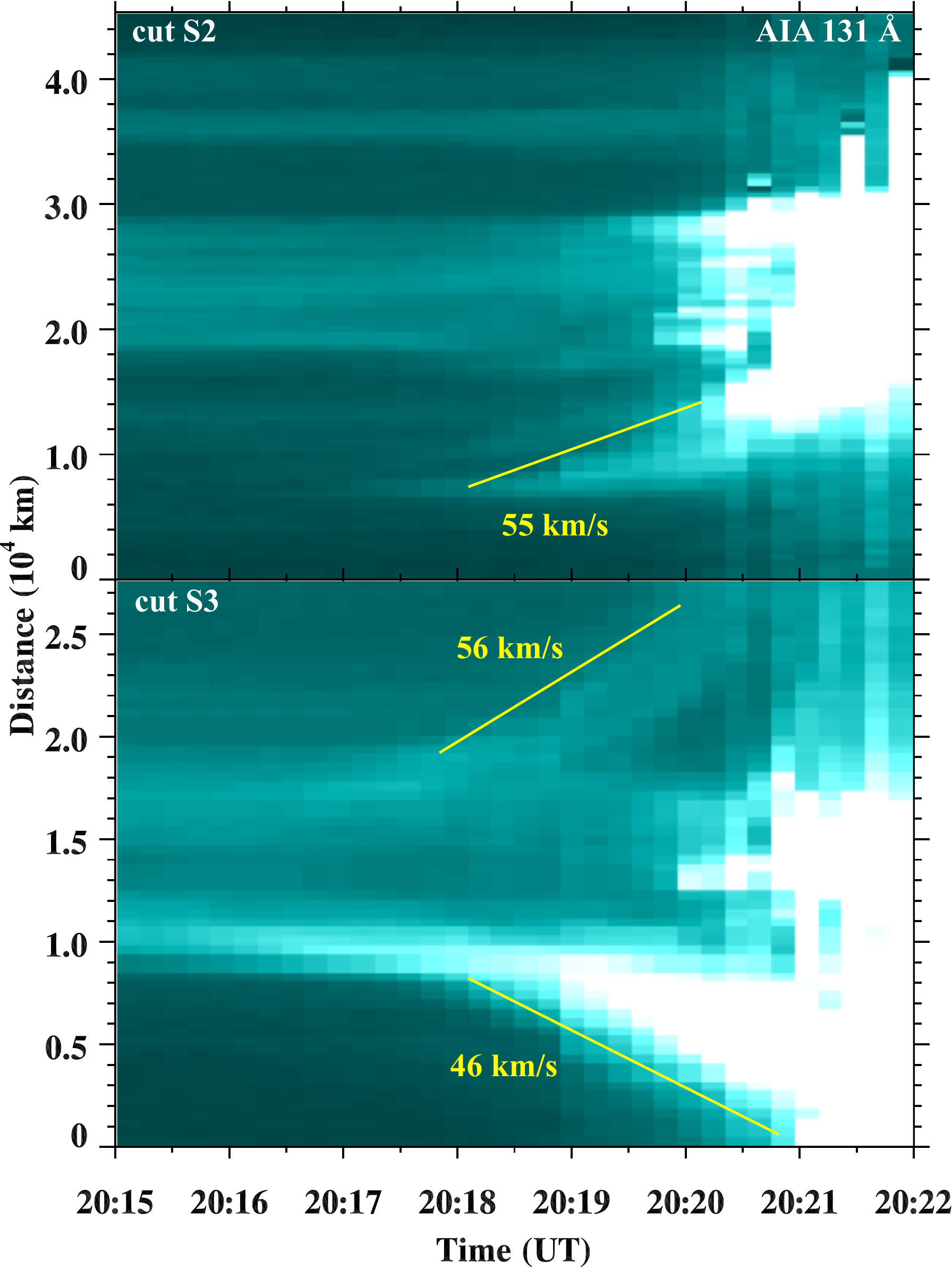}
\includegraphics[width=7cm]{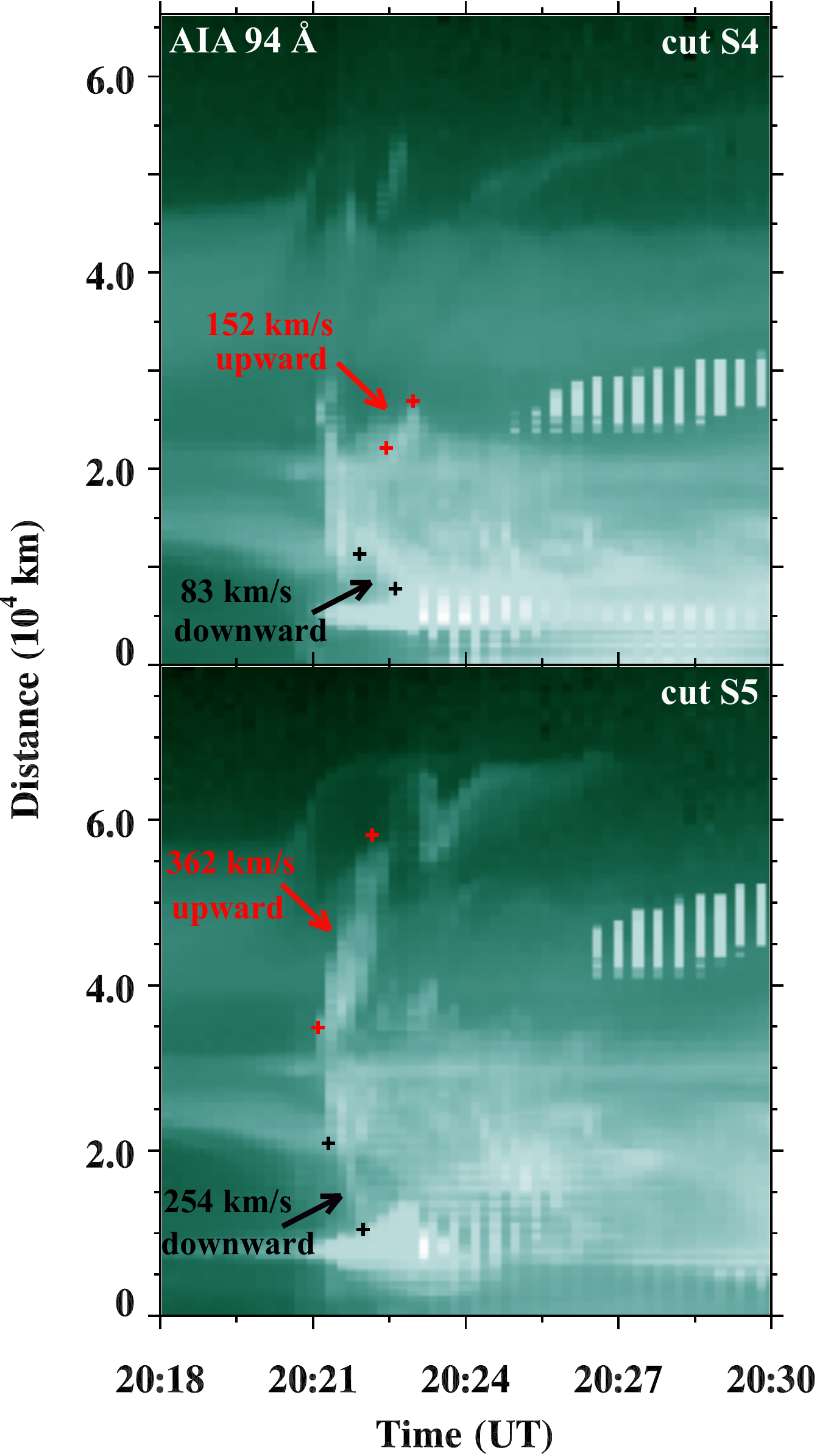}
}
\caption{Space-time intensity plots along slices `S2',`S3',`S4', and `S5' for the AIA 131 (left) and 94 (right) \AA~ images, shown in Figure \ref{aia171}.}
\label{sl}
\end{figure*}

 The top panel of Figure \ref{flux} shows the GOES soft X-ray flux profile (1-8 \AA~ channel) and the distance-time profiles of the upward and downward moving plasmoids (shown in Figures \ref{aia171} and \ref{sl}). The middle panel displays the soft X-ray flux derivative profile. The bottom panel shows 1 sec cadence relative radio flux density (in sfu) profiles at different frequencies (1415, 4995, 8800, and 15400 MHz) observed by the Sagamore Hill radio station \citep{straka1977}. The two stages of energy release is evident during the flare. The first stage energy release (impulsive) was observed during 20:20-20:23 UT. Most of the nonthermal particles are accelerated during the first stage energy release that peaks around 20:21 UT. The 1445 MHz radio flux profile shows quasi-periodic oscillatory behavior. In addition, the radio emission at 4995, 8800, and 15400 MHz is usually related to the gyrosynchrotron emission during the flare. It is likely that the radio emission mechanism at high frequencies ($>$ 2 GHz) is considered due to gyrosynchrotron emission whereas plasma emission below about 2 GHz \citep{dulk1985}. The gradual second energy-release takes place during 20:23 to 20:28 UT.
The timing of the DPSs is indicated by two vertical dotted lines. It is important to note that the DPSs are observed during the first energy release and are well separated from the gyrosynchotron emission produced by the accelerated electrons during the magnetic reconnection.

\subsection{Filament activation}

AR NOAA 11339 was located at N18E49 on 3 November 2011, with the $\beta$$\gamma$ magnetic configuration. This active region produced seven C-class, one M-class and one X-class flare on the same day. According to the Solar Geophysical Data (SGD) report, the X1.9 flare was started at 20:16 UT, peaked at 20:27 UT and ended at 20:32 UT.   
Figure \ref{aia304} displays the selected snapshots of AIA 304 \AA~ channel, which represent the lower solar atmosphere, i.e.,  chromosphere and transition region. These images show the pre-flare activities in and around the active region, where the X-class flare was triggered. About 50 minute prior to the flare, we noticed the activation of a remote-filament (at $\sim$19:30:08 UT) associated with small-scale brightening below it. The filament was located  $\sim$50$\arcsec$ away (toward the north) from the AR (indicated by the arrow). The activated filament slowly rises-up and bends toward the AR after reaching a projected height of $\sim$49 Mm (see image at 19:35:08 UT). AIA 304 \AA~ intensity movie clearly shows the plasma flows along the filament channel from the activation site (direction is shown by arrows). At 19:50:08 UT, we observe plasma flows along `A' and `B' directions toward the AR site.  The filament system starts fading after 20:00 UT, but the plasma flows continued (marked by arrow). The flare was initiated at 20:16 UT, where the plasma flows along the filament channel were dissipated. The flare ribbons are observed at 20:20:08 UT (shown by the arrows). To investigate the photospheric magnetic field configuration at the flare site, we overlaid the HMI magnetogram image contours of positive (white) and negative (black) polarities at the top of AIA 304 \AA~ image at 20:10:08 UT. This image suggests that the filament activation and brightening take place at the location of the weak magnetic flux, where no sunspot was observed in the HMI. 

To explore the coronal magnetic configuration of the AR, we used AIA 131 \AA~ images, which are shown in the bottom panel of Figure \ref{aia304}. The AIA 131 \AA~ channel is sensitive to both cool ($\sim$0.4 MK) and hot ($\sim$11 MK) plasma. The first panel at 19:30:33 UT shows the activation and rising of a remote kinked (indicated by `K') filament. Apparently the kinked filament shows one turn. AIA 131 \AA~ movie reveals the presence of a flux-rope structure (above the flare site) prior to the flare initiation. The flux-rope was visible only in the AIA hot channels (i.e., 131 and 94 \AA), which is marked by `hot FR'. An underlying hot loop is evident at the flare site. The footpoints of the underlying hot loop are indicated by arrows, where the flare ribbons were formed. In the second panel at 19:50:45 UT, we observe the downward motion of the filament apex (`F') near the hot flux-rope (FR) structure, where the flare was triggered. In addition, we also notice the downward motion of the plasma along `B'. We observed underlying hot flare loop at 20:18:33 UT, nearby where  the downward moving plasma along the filament channel disappeared. During the early phase of the X-class flare, we noticed the upward motion of the higher loops (see AIA 131 \AA~ intensity movie, Figure \ref{aia304}) and the downward motion of underlying hot loop. The downward plasma flows along the activated filament channel are likely to have some effect or interaction (i.e., field destabilization) with the ambient magnetic field configuration at the flare site. Although, the present observations are not sufficient to predict the exact triggering process/mechanism of the flare.

To examine the temporal/spatial evolution of the plasma flows towards the AR, we used a slice cut `S1' (AIA 131 \AA~ image  shown in the bottom panel of Figure \ref{aia304}) along the bending filament channel. The space-time intensity plot along with the GOES soft X-ray flux profile (in 1-8 \AA~ wavelength channel) is displayed in Figure \ref{sl131}.
The stack plot reveals the plasma flows along the filament channel till $\sim$20:14 UT (indicated by the arrows).
A vertical red dotted line at $\sim$19:30 UT, indicates the timing of filament activation and associated brightening. We observed two peaks in the flux profile at $\sim$20:22 UT and $\sim$20:27 UT, which reveal the two stages of energy release during the flare. The second peak is more intense, which represents the X-class flare. The timing of plasmoid ejections (observed in EUV) is indicated by two vertical dotted (yellow) lines during the first impulsive energy release.
\begin{figure*}
\centering
{
\includegraphics[width=12cm]{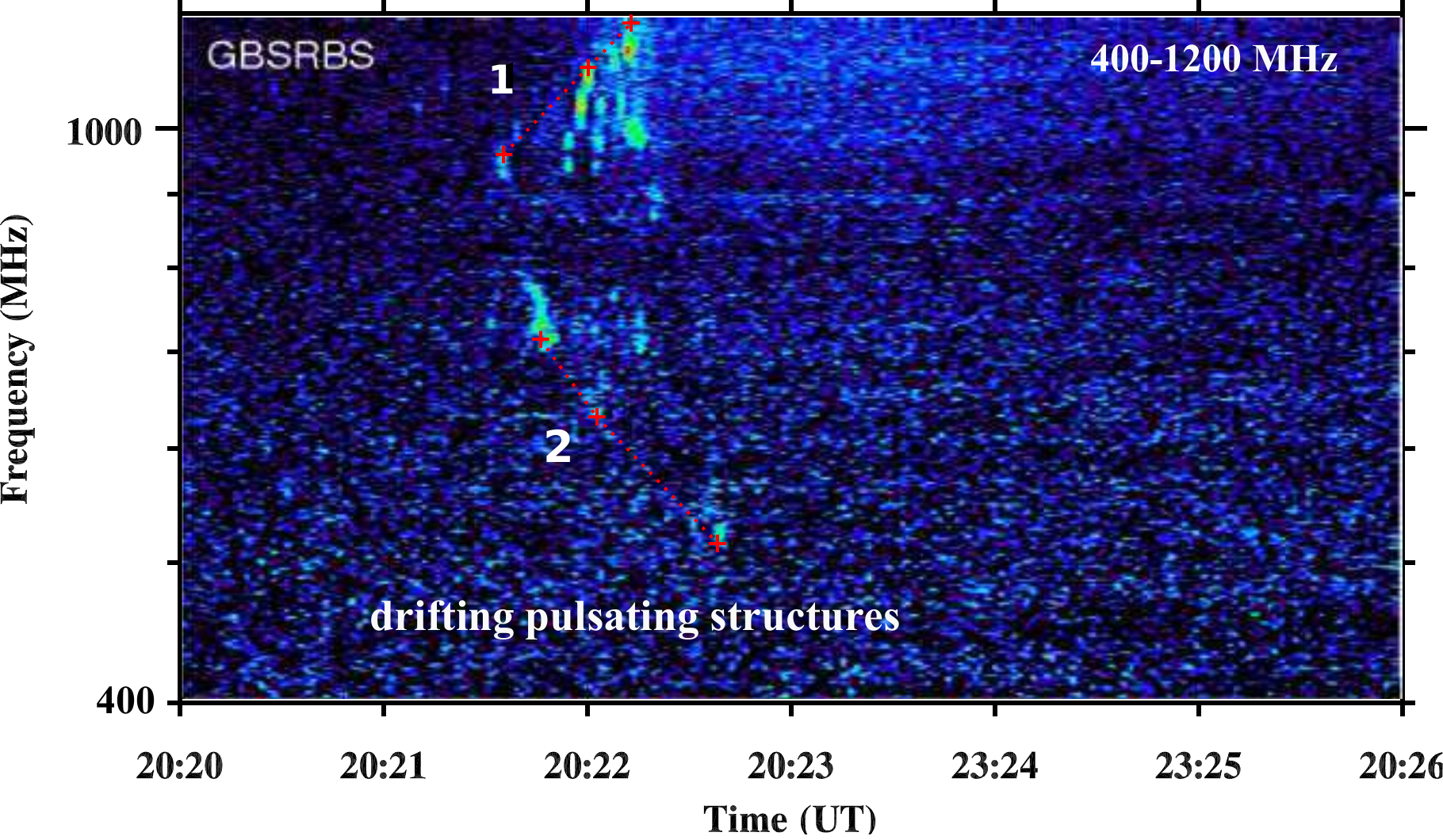}
}
\caption{Dynamic radio spectrum from the GBSRBS showing both the positive (marked by `1') and negative (`2') drifting pulsating structures (DPSs) during the X-class flare on 3 November 2011.}
\label{spectrum}
\end{figure*}

\subsection{Plasmoid ejections}
Figure \ref{aia171} displays the selected snapshots of the AIA 171, 131, and 94 \AA~ during the flare impulsive phase. The first panel (at 20:21:50 UT) shows the ejection of a plasmoid (marked by an arrow) from the flare site, which continues to rise up. Later, we observe multiple plasmoid ejections during the flare impulsive phase (shown in the AIA 171 \AA~ panel at 20:23:27 UT). AIA 171 \AA~ intensity movie shows the interaction and coalescence of multiple plasmoids with each other. The typical thickness of the plasmoids is $\sim$3-4$\arcsec$.
In the bottom panels, we observe a bright underlying flare loop and higher loops in the AIA 131 and 94 \AA~ at 20:18:33 UT. At 20:22:37 UT, multiple plasmoid ejections are observed above the hot underlying flare loop. A vertical structure above the underlying hot loop may be the current sheet (in AIA 131 and 94 \AA~) and the plasmoids are formed/ejected along the sheet structure.
The plasmoids were observed in both hot and cool AIA channels, which suggests the multi-thermal nature of the plasmoids. The plasmoids were impulsively ejected from the current-sheet, and finally diffused into the corona following a deceleration trend. Some of the plasmoids move downward in the corona after reaching a certain height.
To estimate the rising speed of the plasmoid, we tracked the topmost plasmoid (from the flare center) in the AIA 171 \AA~ images shown in the top panels. The distance-time profile of this plasmoid is shown in Figure \ref{flux}. Using the linear fit to the data points, the mean speed of the plasmoid was about 247 km s$^{-1}$. We assume four pixels (2.4$\arcsec$) error in the distance measurement for the visually tracked topmost plasmoid. The estimated error in the mean speed is $\sim$74 km s$^{-1}$. It should be noted that the measured speed is biased by the projection effect and is the lower limit of the true speed of the plasmoid in 3D.

\subsection{Inflows/Outflows}
We used slice cut `S2' and `S3' at AIA 131 \AA~ image to investigate the inflow and outflow structures related to the flare. The directions for slices are chosen after many trials to observe the clear signature of inflow and outflow patterns.
The left panel of Figure \ref{sl} shows the space-time intensity distribution plots for slices `S2' and `S3'. During the initial phase of the flare, we observed the apparent motion ($\sim$55 km s$^{-1}$) of the eastern loop toward the current sheet structure (above the underlying hot loop). At the same time, we observed the upward motion ($\sim$56 km s$^{-1}$) of the higher loops and downward motion ($\sim$46 km s$^{-1}$) of the underlying hot loop system (slice `S3'). These apparent motion of the loop systems may be indirectly linked to the inflow and outflow related to the magnetic reconnection. The outflow speed is quite low during the initial phase of the flare, which is probably the apparent motion of the hot-loop systems.
In a well cited event, \citet{yokoyama2001} found the inflow velocities of 1.0--4.7 km s$^{-1}$ by tracking the motion of patterns above the limb using SOHO/EIT images. Furthermore, \citet{narukage2006} reported the inflow speeds ranging from 2.6--38 km s$^{-1}$ in a statistical study of six flare events. Recently, using the SDO/AIA observations, \citet{savage2012} found the inflow speeds of the order of 100 km s$^{-1}$ and the outflow speeds from $\sim$100 to 1000 km s$^{-1}$. In our observations, the speed of the pattern (i.e., inflow speed) is higher than as reported by \citet{yokoyama2001}, but lower than that of \citet{savage2012}. However, the event studied by \citet{savage2012} was located on the eastern limb, which may account for the higher inflow velocities as the projection effects are minimal.

 We observed the shrinkage of the underlying flare loop ($\sim$46 km s$^{-1}$) about two minutes prior to the radio emission peaks. Simultaneously, we also observed the outward motion of the higher loop ($\sim$56 km s$^{-1}$) and the apparent inflow motion of the loop ($\sim$55 km s$^{-1}$). It seems that the current sheet is formed in between the underlying flare loop and the higher-loop system. Futhermore, the motion of the loops (upwards/downwards) reveal the signature of magnetic reconnection during the early phase of the flare. \citet{sui2003} reported the loop shrinkage (speed$\sim$9 km s$^{-1}$, for 2-4 minutes) before the hard X-ray peak, which was most likely caused by the change of magnetic field configuration as the X-point collapsed into the current sheet. 
Similarly, \citet{li2005} also found the flare loop shrinkage (34 GHz radio images, speed$\sim$13 km s$^{-1}$) in the rising phase of about 9 minutes. Our observational results are consistent with the above reported events, but the shrinkage speed is higher in our case. However, we do not have the radio and hard X-ray imaging observations to study the source motion in more details.

Some of the AIA images in other wavelengths (except 94 \AA~) showed saturation at the flare center during the impulsive phase of the flare. Therefore, it was not possible to observe the downward moving plasmoids in these wavelengths.
To observe the simultaneous upward and downward motion of the plasmoids during the flare impulsive phase, we created the stack plots along slices S4 and S5 using AIA 94 \AA~ images (marked in the bottom-right panel of Figure \ref{aia171}). The start and end points for these slices (in arcsecs) are taken as (i) S4=(-803, 315), (-850, 390) (ii) S5=(-804, 300), (-836, 400).

The right panel of Figure \ref{sl} displays the stack plots along slices S4 (top) and S5 (bottom) during 20:18-20:30 UT. Interestingly, we observed the upward and downward moving plasmoids simultaneously. The position of the upward and downward moving plasmoid is marked by red and black `+' symbols. We estimated the speed of the upward and downward moving plasmoid along slice S5. The initial speed of the upward moving plasmoid is $\sim$362 km s$^{-1}$, whereas the speed of the downward moving plasmoid is $\sim$254 km s$^{-1}$. Note that these speeds are the initial speeds of the plasmoid, which are tracked in the space-time plot. The speed of the upward moving plasmoid is slightly more than the average speed ($\sim$247 km s$^{-1}$) of the topmost plasmoid as reported in the previous section. The calculated speed of the other plasmoids (along slice S4) is $\sim$152 km s$^{-1}$ (upward) and $\sim$83 km  s$^{-1}$ (downward).

The characteristics of the flare (i.e., inflow and outflow) are consistent with the standard flare model (i.e, CSHKP). To determine the reconnection rate from our observations, we consider the outflow speed equal to the Alf\'ven speed, i.e., V$_{o}$=V$_A$ as predicted in the two-dimensional magnetic reconnection theories \citep{priest2000}. Assuming the speed of pattern ($\sim$55 km s$^{-1}$) roughly equivalent to the inflow speed. The reconnection rate is $M_{A}=\frac{V_i}{V_{o}}$$\sim$0.22, for the outflow speed of $\sim$247 km s$^{-1}$, which is consistent with the reconnection rate reported in \citet{takasao2012}. This reveals the fast-reconnection as predicted in \citet{petschek1964}
reconnection model. Several plasma blobs appeared in the current-sheet structure, which collided with each other and ejected from it.
On the basis of the observational findings, we consider that the sheet structure is the current sheet and the plasma blobs are the magnetic islands/plasmoids created by the tearing mode instability as suggested by \citet{takasao2012}.

 \subsection{Drifting Pulsating Structures (DPSs)}
Figure \ref{spectrum} shows the dynamic radio spectrum obtained from the Green Bank Solar Radio Burst Spectrometer (GBSRBS) ranging from 400 to 1200 MHz (decimetric) with 1 s time resolution. The spectrometer is located in a radio-quiet zone at NRAO's Green Bank site, therefore it produces a highly sensitive dynamic spectra at low-noise radio interference. We observed drifting pulsating structures (DPSs) during the flare impulsive phase (20:21:24 UT--20:22:36 UT) for about 1 min duration. The positive and negative DPSs are marked by `1' and `2', respectively. It is interesting to note that at the same time, we observed multiple plasmoid ejections in the AIA images (see Figure \ref{aia171}). Therefore, the DPSs are closely related with the dynamics of multiple plasmoid ejections. The negative DPSs are generally observed during the upward moving plasmoids and can be used to derive the source speed. 

The plasma frequency is related with the local electron density by f=9$\surd$n$_e$ MHz, where n$_e$ is the electron density in m$^{-3}$. The frequency drift rate is defined as

\begin{equation}
\frac{df}{dt}\backsimeq\frac{f}{2H}v
\end{equation}

where H=$\mid$n$_e$/$\frac{dn_e}{dr}$$\mid$ is the inhomogeneous plasma density scale height in the source and v=$\frac{dr}{dt}$ is the speed of the moving emission source \citep{poh2007,wang2012}. Then, the source speed can be estimated by using 
\begin{equation}
v=2\frac{1}{f}\frac{df}{dt}H
\end{equation}
 where $\frac{1}{f}\frac{df}{dt}$ is the relative frequency drift-rate (s$^{-1}$).
 To estimate the drift-rates for the DPSs, we selected few data points at the emission lane (marked by `+' symbol) in the dynamic spectrum and fitted a straight line to get the slopes, i.e., frequency drift rates. The estimated relative frequency drift-rates are 0.0055 and --0.0061 s$^{-1}$, respectively for `1' and `2'. In a different DPSs event, \citet{khan2002} measured the frequency drift rate of --2.8 MHz s$^{-1}$ at a mean frequency of 430 MHz. Our calculated relative negative drift rate is comparable with the value of about --0.0065 s$^{-1}$ in \citet{khan2002}. The positive drift represents the downward moving emission source. From the AIA observations, the speed of upward moving plasmoid is $\sim$247 km s$^{-1}$ (Figures \ref{aia171} and \ref{flux}). Therefore, using equation (2), we obtain the scale height H=2.0$\times$10$^{4}$ km. This value is in agreement with the density scale heights reported by \citet{asc1986} (H$\sim$(2--20)$\times$10$^{3}$ km) by analyzing several type III radio bursts and pulsations in a frequency range of 100--1000 MHz.
 Using above scale height, we obtained the speed of the downward moving radio source  $\sim$224 km s$^{-1}$. Note that these are the mean speed of the upward and downward moving radio sources for the observed DPSs. 
 The speed of the downward moving radio source (from radio observation) is less than the speed of the upward moving source, i.e., plasmoid. The observed drifting-pulsating emissions are likely to be generated by the plasma emission. The positive and negative DPSs implies the plasma densities of 1.13$\times$10$^{10}$ to 1.72$\times$10$^{10}$ cm$^{-3}$ and 3.2$\times$10$^{9}$ to 6.2$\times$10$^{9}$ cm$^{-3}$, respectively.

To determine the speed of the DPS exciter independently from the coronal density model, we use the Newkirk one-fold and two-fold density models \citep{newkirk1961}. But, these models do not provide the reliable heights of the emission source, which lies below the heliocentric distance of 1 R$_\odot$. It means that the emission source is located below the solar surface, which is not correct. Alternatively, we use a barometric isothermal density law for the local density scale height, which is generally used for the decimetric emission \citep{poh2007,poh2008}. In this method, the local electron density (n$_e$) is given by,

\begin{equation}
n_e=N_e exp(-\frac{696000}{H}(1-\frac{1}{R}))
\end{equation}

where N$_e$ is the reference electron density at the base of the corona and R is the estimated heliocentric distance \citep{demoulin2000}. If we assume the reference electron density of 10$^9$ cm$^{-3}$, local scale height of 2$\times$10$^4$ km, and the local electron densities from the DPS. The estimated exciter speed for the positive and negative DPS are 194 and 228 km s$^{-1}$, respectively. However, the speeds are closer to the above estimated speeds, but the source heights are still below the heliocentric distance of 1 R$_{\odot}$ (0.92-0.93 R$_{\odot}$). Therefore, we would like to mention that without proper normalization, density models may not provide the true height of the exciter in the case of the high frequency decimetric bursts (i.e., type II) or drifting pulses \citep{bain2012,cho2013}.


\begin{table*}
\caption{Temperature (MK) and density (cm$^{-3}$) distributions within the selected regions}             
\label{table:1}     
\centering                                      
\begin{tabular}{c c c c c c c c}         
\hline\hline                        
Date/Time(UT) & boxes & log(T$_p$) (MK) & log(EM$_p$) (cm$^{-5}$ K$^{-1}$) & $\sigma_T$ & TEM (cm$^{-5}$) & L &n$_e$ (cm$^{-3}$) \\    
\hline                                   
 03/11/2011 20:18:38& 1 & 6.73  &22.28  &0.149  &8.63$\times$10$^{28}$  &12$\arcsec$  &9.73$\times$10$^9 $                                 \\      
     &2  &6.80  &22.31 &0.149  &1.10$\times$10$^{29}$  &8$\arcsec$ &1.35$\times$10$^{10}$                 \\
     &3  &6.92  &22.39 &0.149  &1.71$\times$10$^{29}$  &8.5$\arcsec$   &1.63$\times$10$^{10}$  \\ \hline
03/11/2011 20:22:38&1  &6.86  &22.01  &0.170  &7.30$\times$10$^{28}$  &13$\arcsec$   &8.65$\times$10$^9$                            \\
     &2  &6.29  &22.41  &0.170  &4.92$\times$10$^{28}$  &9$\arcsec$   &8.54$\times$10$^9$                             \\
      &3  &6.54  &22.10  &0.170  &4.24$\times$10$^{28}$  &6.4$\arcsec$   &9.39$\times$10$^9$                             \\
       &4  &6.22  &22.58  &0.170  &6.19$\times$10$^{28}$  &5$\arcsec$   &1.28$\times$10$^{10}$                            \\
        &5  &6.24  &22.42  &0.170  &4.45$\times$10$^{28}$  &3$\arcsec$   & 1.40$\times$10$^{10}$                            \\
         &6  &6.28  &23.09  &0.170  &2.28$\times$10$^{29}$  &4$\arcsec$   &2.76$\times$10$^{10}$   \\ \hline
18/08/2010 05:10:50& 1 & 6.39  &21.90  &0.175  &2.01$\times$10$^{28}$  &2$\arcsec$  &1.15$\times$10$^{10} $                                 \\      
     &2  &6.30  &22.13 &0.175  &2.76$\times$10$^{28}$  &6$\arcsec$ &7.83$\times$10$^{9}$                 \\
     &3  &6.18  &22.03 &0.175  &1.62$\times$10$^{28}$  &3$\arcsec$   &8.5$\times$10$^{9}$  \\

\hline                                          
\end{tabular}
\end{table*}

\subsection{Differential Emission Measure (DEM) analysis of the active region} 
\begin{figure*}
\centering{
\includegraphics[width=6cm]{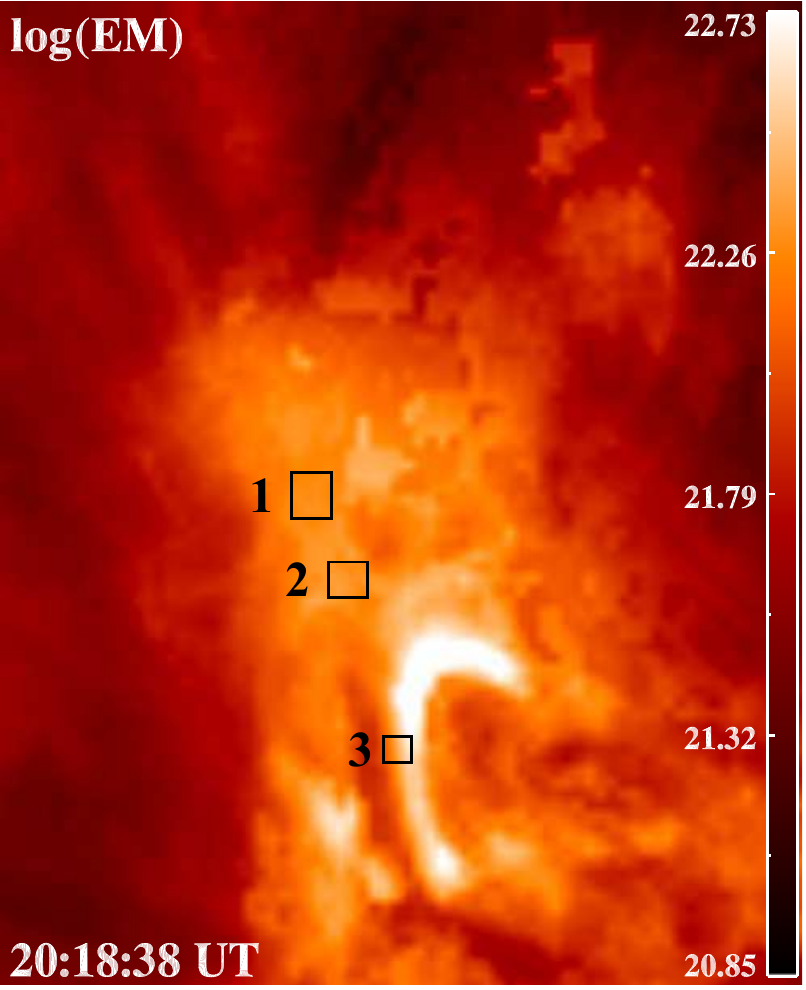}
\includegraphics[width=6cm]{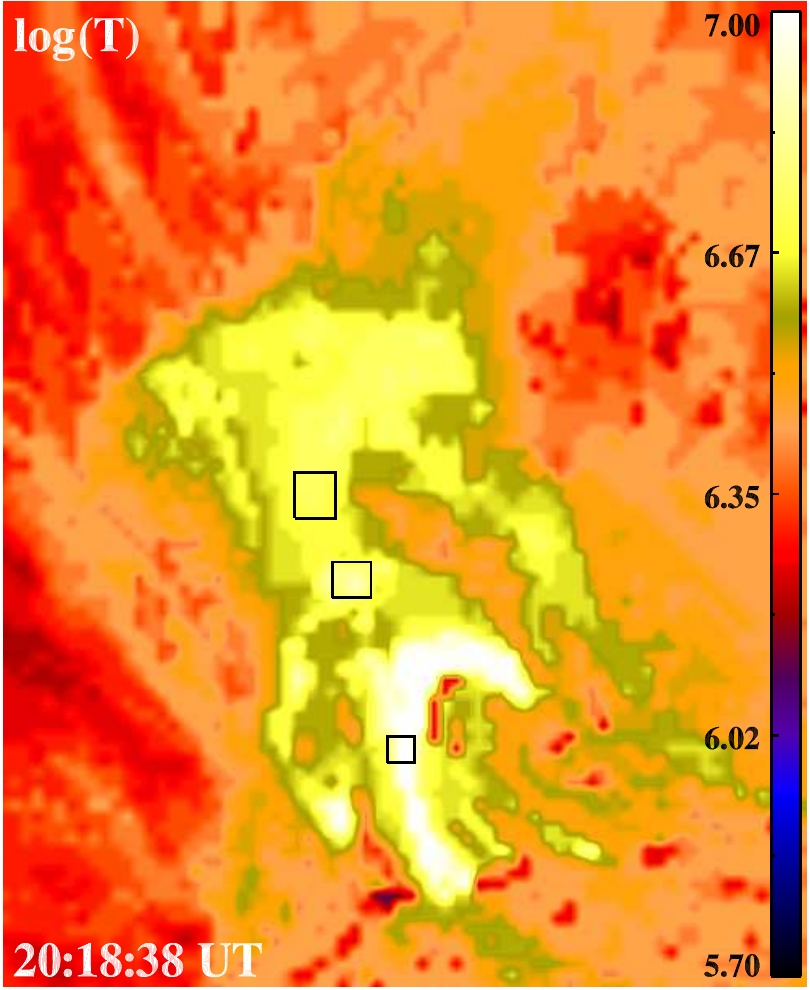}
}

\centering{
\includegraphics[width=6cm]{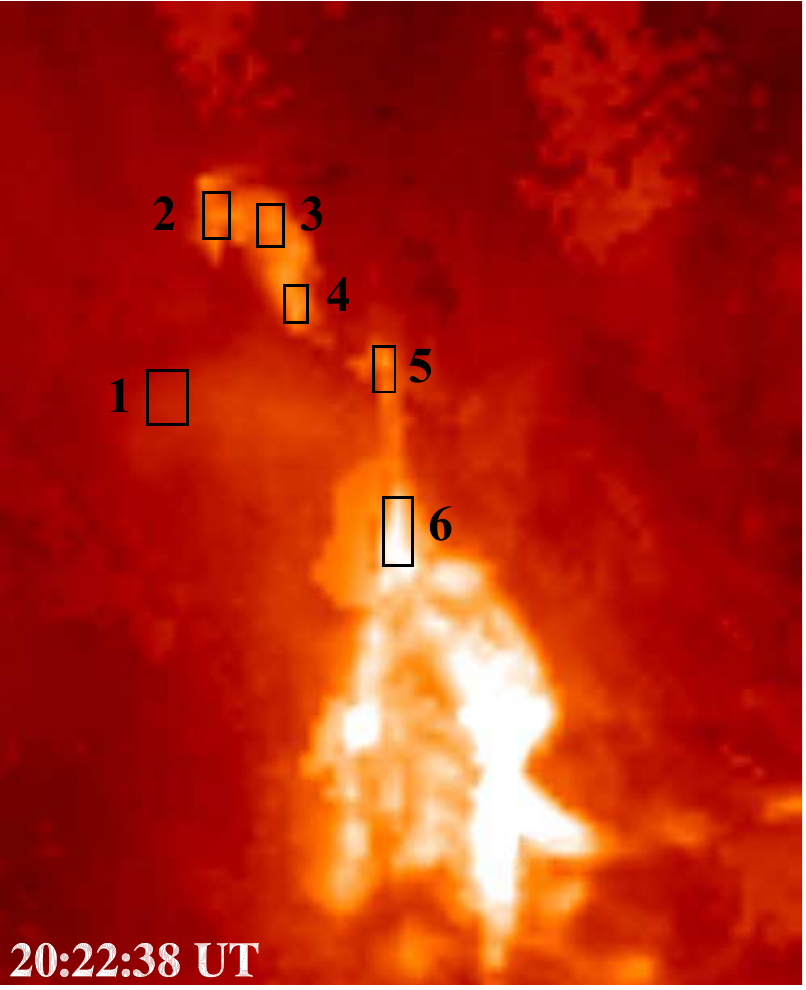}
\includegraphics[width=6cm]{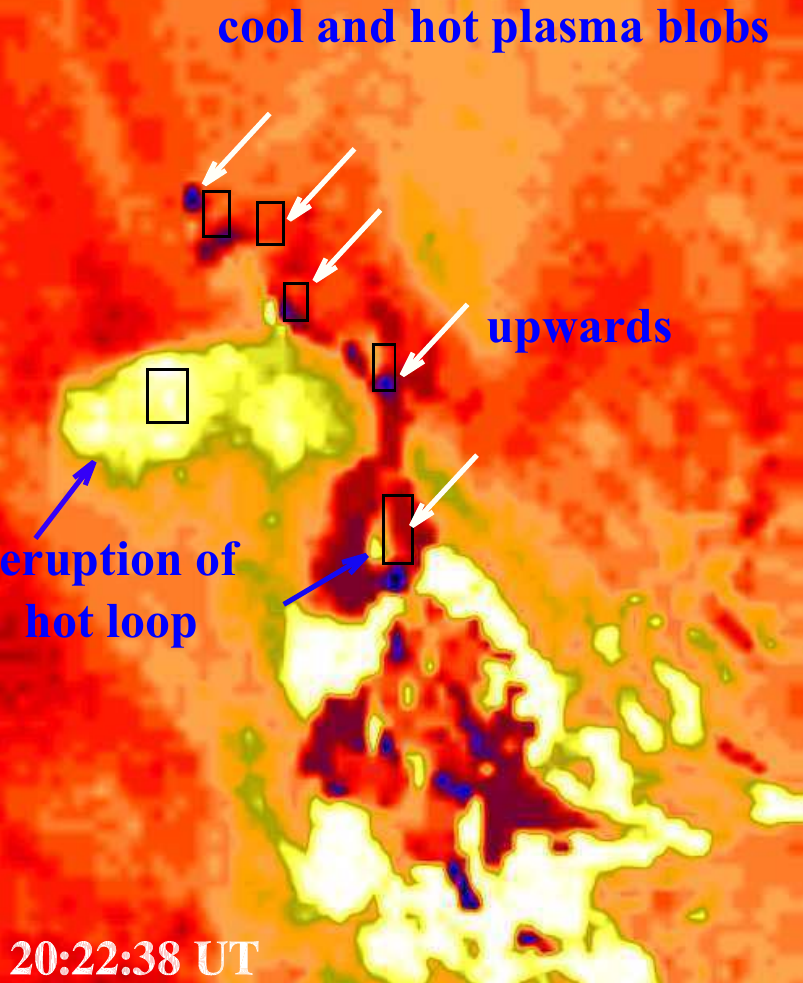}
}

\caption{Peak emission measure (cm$^{-5}$ K$^{-1}$) and temperature (MK) maps for the active region during the flare on 3 November 2011. The size of each image is 85$\arcsec$$\times$104$\arcsec$.}
\label{dem1}
\end{figure*}

To determine the peak temperature and emission measure of the active region, we utilized AIA images in six coronal filters (i.e., 94, 171, 131, 211, 335, and 193 \AA). We used the SSWIDL code developed by \citet{asc2011} for this purpose. The coalignment of AIA images at six coronal EUV wavelengths is carried out by using the limb fitting method, with an accuracy of $<$1  pixel. The code fits a DEM solution in each pixel (or macro-pixel), which can be parametrized by a single Gaussian function that has three free parameters: the peak emission measure (EM$_p$), the peak temperature (T$_p$), and the temperature width sigma ($\sigma_T$). The peak emission measure (cm$^{-5}$ K$^{-1}$) and temperature (MK) maps (during the flare and plasmoids eruption) are shown in Figure \ref{dem1}.
Furthermore, to estimate the density of the hot coronal loop, the current sheet structure, and the multiple plasmoids, we selected small boxes that are shown in the top and bottom panels of Figure \ref{dem1}. First of all, we calculated the average temperature, EM and $\sigma_T$ in the selected regions. Using these values, we estimated the total emission measure (TEM) in the selected regions using formula, $\int \! DEM(T) \, \mathrm{d} T$.
Using the values of total EM (cm$^{-5}$), we estimated the densities of the selected structures in the active region, assuming that the depth of the structure along the line of sight is nearly equal to its width \citep{cheng2012}. If `L' is the width of the structure, then the density (n$_e$) of the structure is calculated by using the relation $n_e=\sqrt{\frac{EM}{L}}$ (assuming filling factor $\approx$1). The width of structures are marked in the AIA 94 \AA~ images (Figure \ref{aia171}, bottom panels). All the estimated values are summarized and listed in Table 1.

\begin{figure*}
\centering{
\includegraphics[width=6cm]{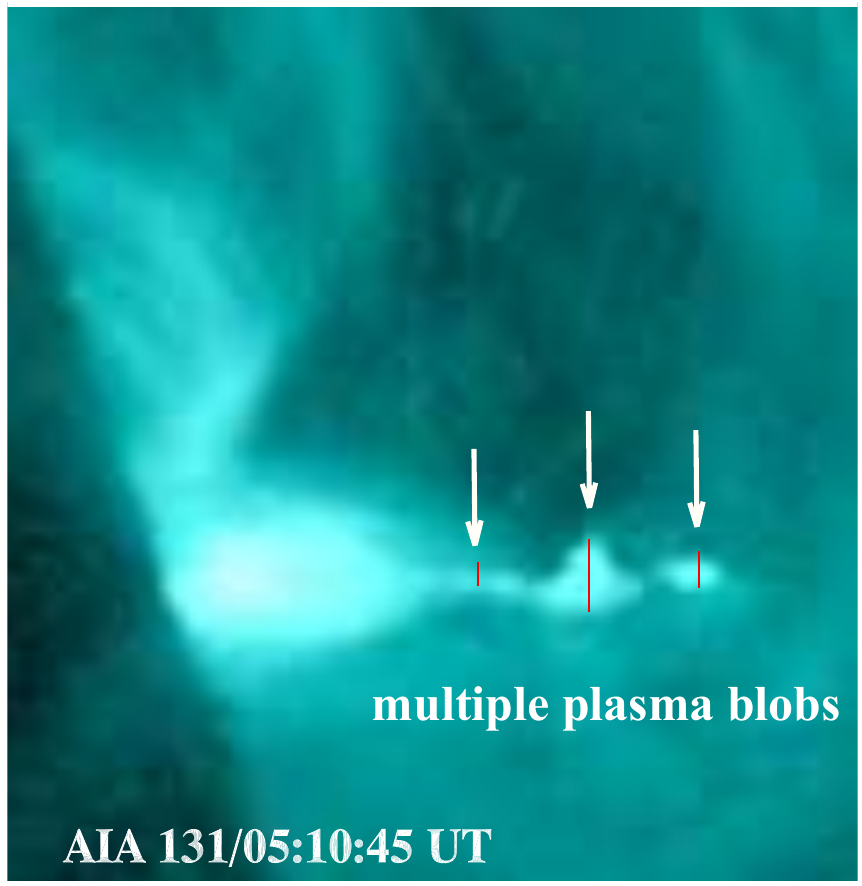}
\includegraphics[width=6cm]{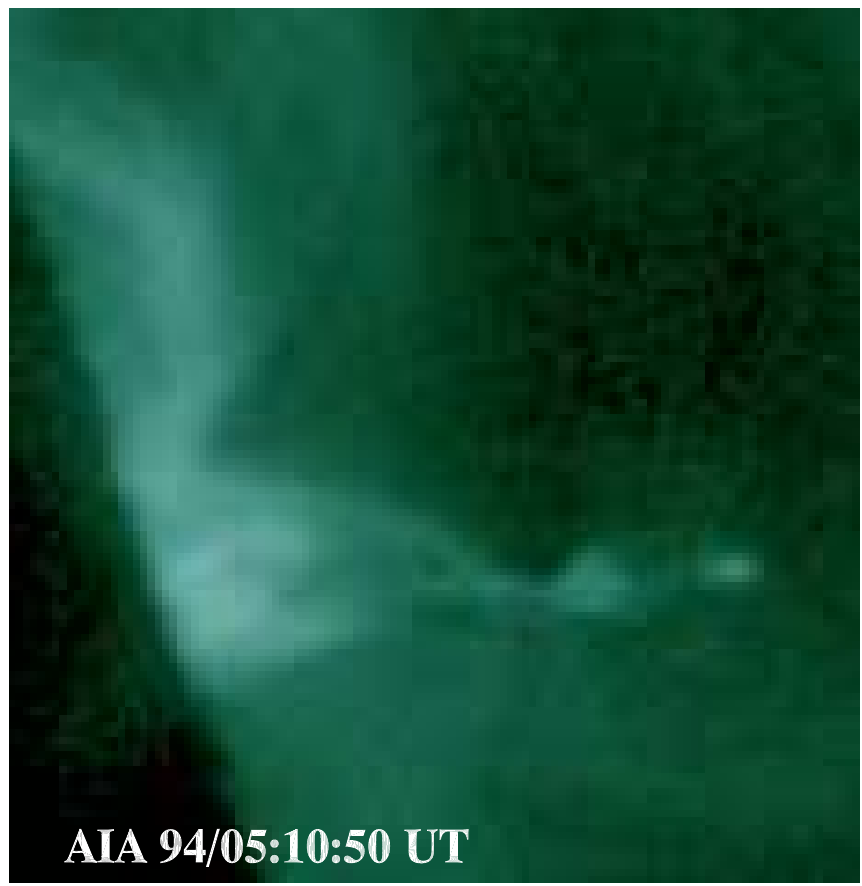}
}

\centering{
\includegraphics[width=6cm]{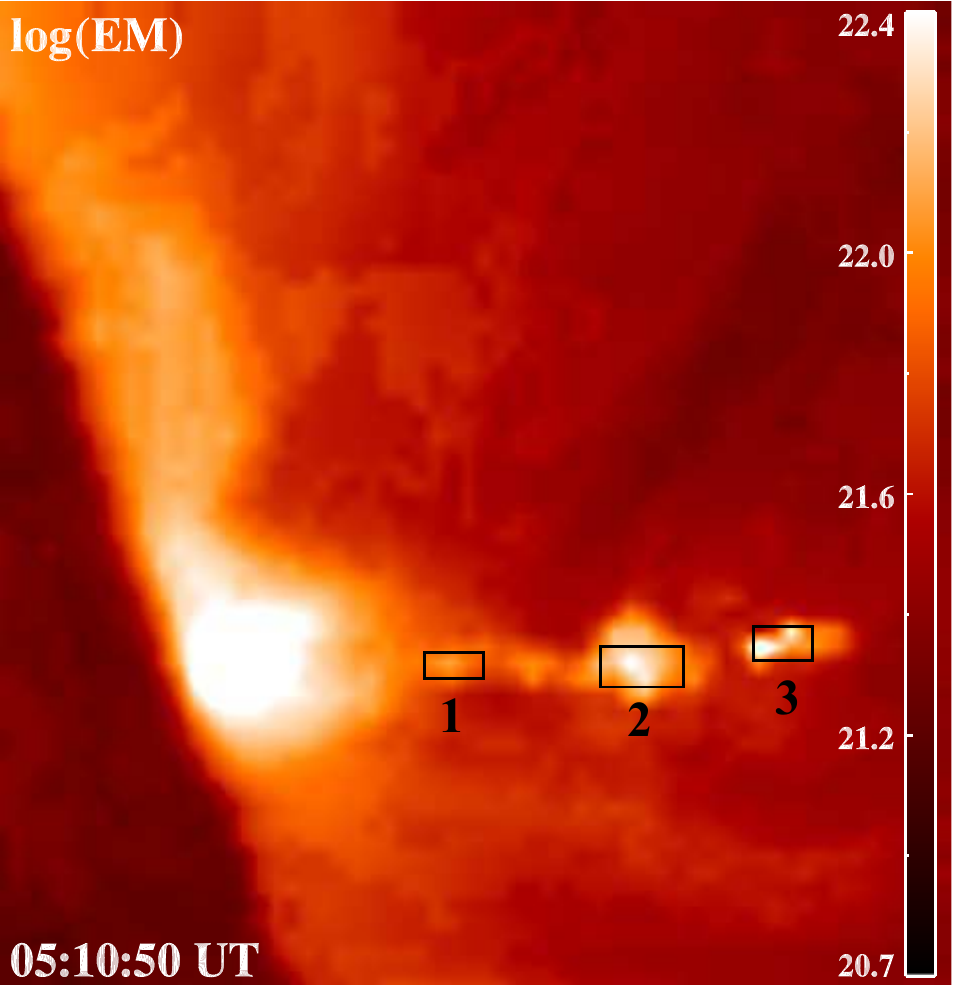}
\includegraphics[width=6cm]{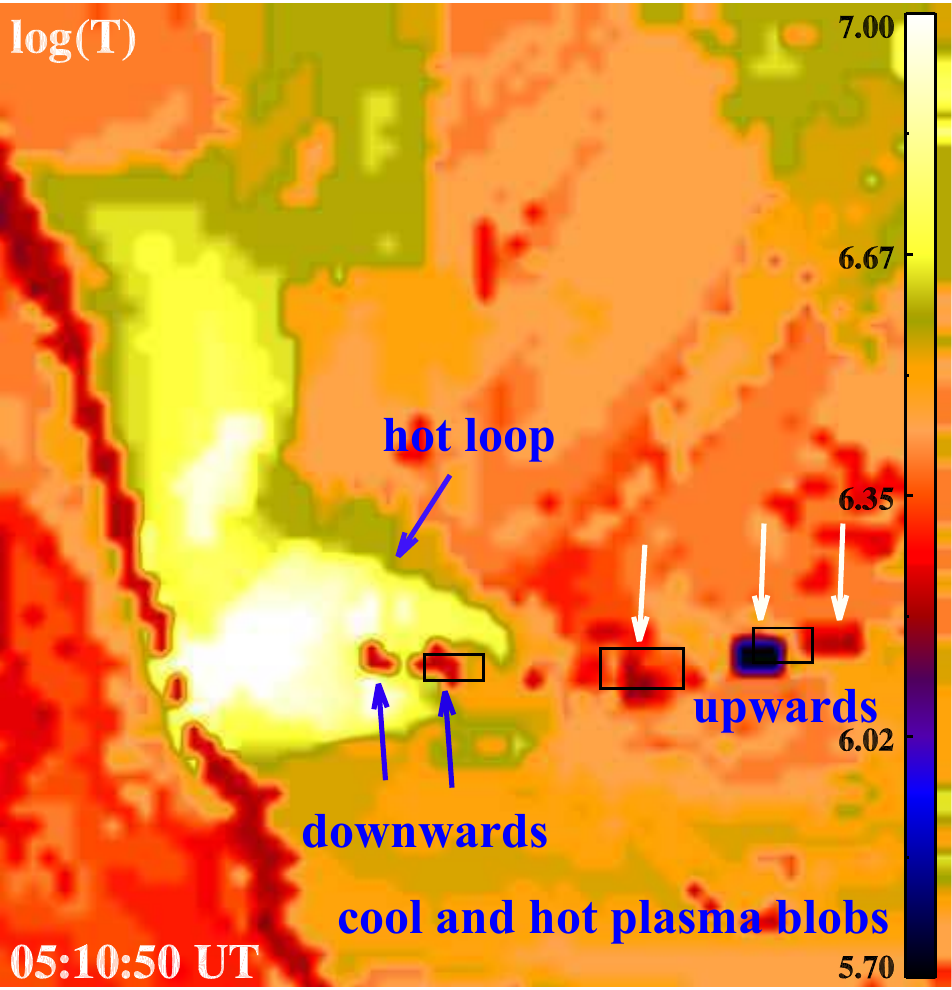}
}

\caption{Top: AIA 131 and 94 \AA~ images showing the ejection of multiple plasmoids. The red lines in the first panel indicate the approximate thickness of the plasmoids. Bottom: Peak emission measure (cm$^{-5}$ K$^{-1}$) and temperature (MK) maps for the active region during C4.5 flare occurred on 18 August 2010. The size of each image is 81$\arcsec$$\times$84$\arcsec$.}
\label{dem2}
\end{figure*}

In Figure \ref{dem1}, we displayed the peak emission measure and the peak temperature maps at 20:18:38 UT (top) and 20:22:38 UT (bottom). The top panels show a dense hot underlying loop, a possible current-sheet structure and a higher loop system above it. The estimated mean temperature within the selected regions (indicated by 1, 2, 3) are 5.37, 6.30, and 8.31 MK, respectively. This reveals a quite high temperature of these structures/loops, which are observed only in AIA 131 and 94 \AA~ channels. This indicates the presence of a hot flux-rope before the triggering of the flare. The computed mean density within these regions are $\sim$$9.73\times10^9$, $\sim$$1.35\times10^{10}$ and $\sim$$1.63\times10^{10}$ cm$^{-3}$, respectively. The bottom panels illustrate the ejection of multiple plasmoids along the current sheet structure. The ejection of hot plasma possibly from the current sheet structure is evident, which is marked by `1'. The peak temperature and density of the ejected hot plasma are $\sim$7.2 MK and $\sim$8.65$\times$10$^{9}$ cm$^{-3}$, respectively . The rest boxes from `2' to `6' show the multiple plasmoid ejections along the current sheet.
It is interesting to note that the cool and hot plasma blobs are detected along the current-sheet structure. The peak temperature of the plasma blobs varies from $\sim$1.6 to 3.4 MK, whereas the density varies from $\sim$8.54$\times$10$^{9}$ to $\sim$2.76$\times$10$^{10}$ cm$^{-3}$. Although we could observe the downward moving plasmoids in the AIA 94 \AA~ images, it was not possible to track these plasmoids in the temperature maps due to the image saturation in other AIA channels.

A similar type of bidirectional multiple plasmoid ejections associated with inflows is recently reported by \citet{takasao2012} in a C4.5 limb flare occurred in AR NOAA 11099 on 18 August 2010. The plasma blobs were observed both in hot and cool AIA channels during the impulsive phase of the flare similar to our event. The top panels of Figure \ref{dem2} (i.e., AIA 131 and 94 \AA~ images) show the underlying hot loop and the ejection of multiple plasmoids above it, along the possible current sheet structure. To compare the characteristics of the plasma blobs in both events, we generate the peak temperature and emission maps for AR NOAA 11099, which is shown in the bottom panel of Figure \ref{dem2}. The peak temperature and density of the plasma blobs in regions 1, 2 and 3 varies from $\sim$1.5 to 2.4 MK and $\sim$7.83$\times$10$^{9}$ to 1.15$\times$10$^{10}$ cm$^{-3}$, respectively. In 18 August 2010 flare, a prominence eruption also takes place nearby the flare site, similar to our event. Later, we observe the flare and ejection of multiple plasmoids. The generation of tearing instability and the ejection of multiple plasmoids are quite similar in both of the events. Although, the radio observations for 18 august 2010 event are not available, otherwise it would have been useful to compare the dynamic radio spectrum for both events.

Interestingly, we observe the multi-temperature plasmoids ($\sim$1.6--3.4 MK) formed by the tearing of current sheet during the magnetic reconnection. The observed temperature of multiple X-ray plasmoids (in a different event) associated with the individual peak of hard X-ray emission was $\sim$10  MK \citep{nishizuka2010}. The estimated temperature and density of a X-ray plasmoid observed in 5 October 1992 flare, was $\sim$6--13 MK and $\sim$8--15$\times$10$^{9}$ cm$^{-3}$, respectively \citep{ohyama1998}. In our case, the observed temperature of the plasmoids is lower than that of reported in the previous studies. However, the density of the plasmoids is comparable. In addition, the temperature of the plasmoid was lower than that of the region in between the plasmoid and the underlying flare loop, which is consistent with the magnetic reconnection model \citep{yokoyama1997,yokoyama1998}.
 The formation of multi-temperature plasmoids during the tearing mode instability is not clear (observed here for the first time) and further investigations are needed.

\section{Discussion and conclusion}

We studied multiwavelength observations of the X-class flare which occurred on 3 November 2011. The associated energy-release/brightening below an activated filament induced the plasma flows along the filament channel.  
Furthermore, the downward moving filament apex towards and close to the AR possibly destabilize the magnetic field configuration at the flare site leading to the flare onset. The plasma flows along the filament channel slow down about 15-20 min prior to the flare occurrence. The downward moving plasma flows are likely to be responsible for the onset of inflows needed for the flare initiation. As shown in the numerical simulation by \citet{shen2011}, an initial perturbation is applied to the current sheet to induce the inflow. The AIA images reveal an underlying hot loop before the flare impulsive phase and the formation of a current sheet possibly above the underlying hot loop. Generation of the tearing mode instability in the current sheet leads to the formation and coalescence of multiple plasmoids \citep{furth1963,shibata2001}. The radio observations confirm both positive and negative DPSs simultaneously during 20:21:24 UT-20:22:36 UT. This suggests the signature of electron acceleration associated with the multiple bidirectional plasmoid ejections (observed in the AIA images) from the reconnection site. The interaction and coalescence of a series of plasmoids leads to the effective particle acceleration and associated plasma heating \citep{karl2007}.

Recently, the AIA observations revealed the formation of a hot flux-rope structure prior to and during the solar eruption \citep{cheng2011,zhang2012}, which was considered as a driver of the solar eruption. In the solar flare event observed on 3 November 2010, a hot vertical current sheet was formed below an erupting flux-rope structure (observed in AIA hot channels, i.e., 131 and 94 \AA) \citep{reeves2011,cheng2011,kumar2013c,hannah2013}. \citet{pats2012} also reported the formation of a hot flux-rope prior to an eruption and its subsequent destabilization leads to the onset of a flare/CME.
Therefore, in our event, hot structure above the underlying flare loop (observed in the 131 and 94 \AA) is the flux-rope structure, which is formed prior to the flare initiation. A current sheet is probably formed in between the hot flux-rope and underlying flare loop. As discussed by \citet{asc2004}, a resistive instability is generated in the current sheet, when the driving forces of the inflow exceed the opposing Lorentz force. These driving forces are produced by the sheared magnetic fields leading to the onset of tearing mode instability \citep{furth1963}.
 
\begin{figure}
\centering{
\includegraphics[width=6cm]{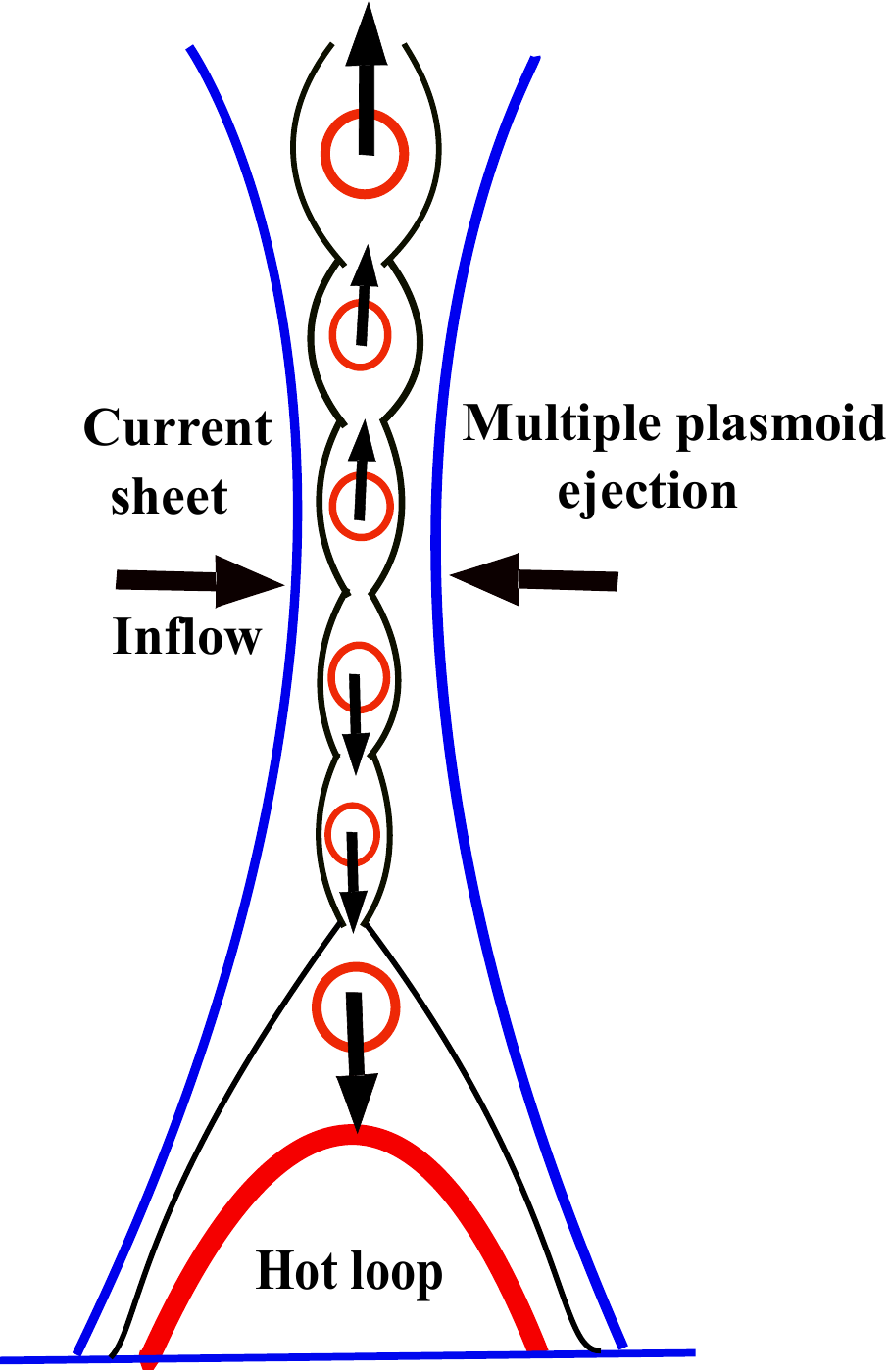}
}
\caption{Schematic cartoon showing the bidirectional plasmoid ejections along the current sheet structure during magnetic reconnection.}
\label{cartoon}
\end{figure}

\citet{shen2011} have performed the resistive MHD simulations to study the internal structure of the current sheets. In their simulation, reconnection-rate becomes faster when the magnetic islands or plasma blobs are formed and the plasmoids move in both directions (up and down) along the current sheet, whose speeds ranging from 147--242 km s$^{-1}$ for upward and 89--159 km s$^{-1}$ for downward moving plasmoids. In our case, the mean speed of the upward and downward moving plasmoids are $\sim$247 and $\sim$224 km s$^{-1}$, respectively. The speed of the downward moving plasmoids is less than the upward moving plsamoids. This may be because the closed flare loops prevent blobs to move faster in the downward/sunward direction \citep{shen2011}. Following Shibata and Karlick\'y models, we draw a schematic cartoon (Figure \ref{cartoon}) to explain the scenario of the event. An underlying flare-loop, inflows, and the bidirectional multiple plasmoid ejections along the current sheet structure are shown.

Using the AIA observations of a limb flare, \citet{takasao2012} observed the signature of inflows and outflows associated with bidirectional plasmoid ejections. The typical size of the plasmoid in their event was about 2-3$\arcsec$ and the velocities of upward and downward ejection were 220-460 km s$^{-1}$ and 250-280 km s$^{-1}$, respectively. The shape, sizes and speed of the plasmoids (thickness about 3-4$\arcsec$) in our event are nearly consistent with their results. \citet{takasao2012} could observe the downward moving plasmoids due to limb event (negligible projection effect). 
Similarly, we also observed the downward motion of the plasmoids in the AIA 94 \AA~ images and the downward motion of the plasmoids is confirmed by the positive DPSs observed in the dynamic radio spectrum. 
Multiple plasmoids/magnetic islands are created probably by tearing mode instability in the current sheet, which collided with each other and were ejected from it. The plasmoids were visible in AIA hot (131 and 94 \AA) and cool (304 \AA) channels, suggesting their multi-thermal nature. Very likely the plasmoids were heated by the coalescence process \citep{kliem2000}. The plasmoid ejections and DPSs are observed during the first impulsive energy release. These plasmoids can induce strong inflow along with the enhancement of the reconnection rate \citep{shibata2001}. Therefore, the ejection of multiple plasmoids play an important role in triggering the second energy release.

\citet{ning2007} also observed the positive DPSs during a flare on 18 March 2003 and suggested that these are probably caused by the downward moving plasmoid in the corona.  Numerical simulation results also show that the downward motion of the plasmoid is possible in current-sheet during a bursty magnetic reconnection \citep{karl2007,shen2011}. Our observational finding of simultaneous upward and downward moving multiple plasmoids supports their interpretation.
 As discussed in the numerical simulation by \citet{karlicky2010}, the pulses of electromagnetic emission are generated at a location between two interacting plasmoids just before the coalescence of two plasmoids into a larger one. They found that the Langmuir waves accumulate in the interacting plasmoids at the locations of superthermal electrons \citep{drake2005,karlicky2007}. On the other hand, the electromagnetic waves appear
at the boundaries of the plasmoids and then move outwards, where they mutually interfere and generate the short-period pulsations. Using the particle in cell (PIC) model with periodic boundary conditions, \citet{oka2010} also found the efficient electron acceleration by multi-island coalescence process.
 Therefore, we expect that the electron acceleration takes place in between the plasmoids during the coalescence/merging of the plasmoids as shown in the model of \citet{karl2004}. Moreover, densities of the plasmoid regions derived with the AIA images, are consistent with the densities corresponding to the positive and negative DPSs.  Thus, our observations are in agreement with the suggestions/interpretations given by \citet{karlicky2010} and \citet{karlicky2011}.

However, in the recent numerical simulations \citep{barta2008,shen2011}, both the negative and positive DPSs have been suggested to be associated with the upward and downward moving plasmoids respectively, but the simultaneous oppositely directed DPSs in radio, associated with a series of plasmoids in EUV, have not been reported earlier.

In conclusion, we reported the simultaneous radio and EUV observations of the multiple plasmoid ejections that moved bidirectionally along the current sheet structure during magnetic reconnection. Furthermore, the high-resolution observations (from Hinode and SDO) of similar type of events combined with radio observations, will be helpful to understand the characteristics of these plasma blobs and associated particle acceleration in more details. 

\begin{acknowledgements}
We would like to thank the anonymous referee for his/her constructive comments/suggestions and help in the interpretation of the data, which
improved the manuscript considerably. 
SDO is a mission for NASA's Living With a Star (LWS) program. National Radio Astronomy Observatory
(NRAO) is operated for the NSF by Associated Universities, Inc.,
under a cooperative agreement. PK thanks P.K. Manoharan and D.E. Innes for several helpful discussions/suggestions. This work has been supported by the ``Development of Korea Space Weather Center" project of KASI, and the KASI basic research fund.
 \end{acknowledgements}
\bibliographystyle{aa}
\bibliography{reference}
\end{document}